\newif\ifuseprd
\newif\iftoomuchdetail
\newif\ifeprint

\useprdtrue % (un)comment if you do/don't prefer revtex4
\toomuchdetailfalse
\eprinttrue
\newif\ifrr
\rrfalse
\newcommand\RR{{\ifrr{RR}\else{Ramond-Ramond (RR)}\fi\global\rrtrue}}

\newif\ifns
\nsfalse

\ifuseprd 
\documentclass[\ifeprint preprint,\fi
tightenlines,floats,prd,eqsecnum,nobibnotes%
,nofootinbib,aps]{revtex4}
\else
\documentclass[final,12pt,letterpaper]{JHEP3}
\fi %useprd
\usepackage{amsmath,amssymb}
\usepackage{pifont}

\allowdisplaybreaks[3]
\newcommand\registered{{\Pisymbol{psy}{226}}}% 210 also works
\newcommand\Mathematica{{Mathematica$^{\text{\registered}}$}}

\newcommand\skipthis[1]{{}}
\newcounter{saveequation}
\newcounter{detailnum}
\newcommand\savetheequation{\theequation}
\newcommand\detailtheequation{%
         D\arabic{detailnum}:\thesection.\arabic{equation}}
\newenvironment{detail}{\iftoomuchdetail\sf% no let's set up equations
         \setcounter{saveequation}{\value{equation}}%
         \setcounter{equation}{0}\addtocounter{detailnum}{1}%
         \renewcommand\theequation\detailtheequation%
         \fi}{% unset up equations
     \iftoomuchdetail%
     \ifnum\value{equation}=0\addtocounter{detailnum}{-1}\fi%
     \setcounter{equation}{\value{saveequation}}%
     \renewcommand\theequation\savetheequation%
     \fi%
     }
\makeatletter
\def\@strike{\relax\leavevmode
  \ifmmode
    \expandafter\mathpalette\expandafter\math@strike
  \else
    \expandafter\make@strike
  \fi}
\def\math@strike#1#2{%
  \setbox\z@\hbox{$\m@th#1{#2}$}\fin@strike}
\def\make@strike#1{%
  \setbox\z@\hbox{\color@begingroup#1\color@endgroup}\fin@strike}
\def\fin@strike{%
  \@tempdima\dp\z@
  \@tempdimb\ht\z@
  \lower\@tempdima\hbox{\strike@start}%
  \box\z@
  \raise\@tempdimb\hbox{\strike@end}}
\def\strike@start{\special{ps: %
    currentpoint /starty exch def /startx exch def}}
\def\strike@end{
}
\newcommand\fs{\protect\@strike}

\let\oldAE\AE
\renewcommand\AE{{\ifmmode{\text{\it\oldAE}}\else{\oldAE}\fi}}

\newlength{\parboxlen}

\newlength{\notlen}

\newcommand\mathone{{\rlap{\kern .25em l}1}}
\newcommand\one{{\ifmmode{\text{\mathone}}\else{\mathone}\fi}}
\newcommand\proj{{\ensuremath{{{\mathbb P}}}}}
\newcommand\ct[1]{{\ifeprint\ifuseprd{\em{#1}},\else{\sf {#1}},\fi\fi}}
\newcommand\bt[1]{{\em {#1}},}

\newenvironment{spmatrix}{\left(\begin{smallmatrix}}{\end{smallmatrix}\right)}

\chardef\til=`~

\newcommand\half{{\ensuremath{\frac{1}{2}}}}

\newcommand\p{\ensuremath{\partial}}
\newcommand\evalat[2]{\ensuremath{\left.{#1}\right|_{#2}}}

\newcommand\norm[1]{\ensuremath{\left\|{#1}\right\|}}

\newcommand\transpose{{\ensuremath{\text{\sf T}}}}
\newcommand\field[1]{{\ensuremath{\mathbb{{#1}}}}}
\newcommand\ZZ{{\field{Z}}}

\newcommand\ZR{{\field{R}}}

\newcommand\anti[2]{\ensuremath{\left\{{#1},{#2}\right\}}}
\newcommand\com[2]{\ensuremath{\left[{#1},{#2}\right]}}
\DeclareMathOperator{\Tr}{Tr}

\newcommand\lie[2]{\ensuremath{\pounds_{{#1}} {#2}}}
\newcommand\apr{{\ensuremath{{\alpha'}}}}

\DeclareMathOperator{\SO}{SO}
\DeclareMathOperator{\SU}{SU}

\providecommand\putabstract[1]{\ifuseprd\begin{abstract} {#1} \end{abstract}%
                           \else \abstract{{#1}} \fi}
\providecommand\plb[3]{{Phys.\ Lett.\ B {\bf {#1}}, {#3} ({#2})}}
\providecommand\npb[3]{{Nucl.\ Phys.\ {\bf B{#1}}, {#3} ({#2})}}
\providecommand\jhep[3]{{J.\ High Energy Phys.\ {\bf #1}, {#3} ({#2})}}

\providecommand\mpla[3]{{\ifuseprd\else\begingroup\em\fi Mod.\ Phys.\ Lett.\ %
     \ifuseprd\else\endgroup\fi {\bf A{#1}}\ifuseprd, {#3} ({#2})\else%
     \ ({#2}) {#3}\fi}}
\providecommand\lnp[3]{{\ifuseprd\else\begingroup\em\fi Lect.\ Notes\ Phys.\ %
     \ifuseprd\else\endgroup\fi {\bf {#1}}\ifuseprd, {#3} ({#2})\else%
     \ ({#2}) {#3}\fi}}
\newcommand\citeprd[3]{{\ifuseprd{Phys.\ Rev.\ D {\bf {#1}}, {#3} ({#2})}%
                        \else{\prd{#1}{#2}{#3}}\fi}}

\providecommand\hepth[1]{{\ifuseprd{\eprint{{\ifeprint\tt\fi hep-th/#1}}}%
                \else{\tt hep-th/{#1}}\fi}}
\newcommand\web[1]{{\ifuseprd{\hbox{\url{#1}}}\else{\tt \hbox{{#1}}}\fi}}
\newcommand\phepth[1]{{\ifuseprd\else\tt\fi [\hepth{#1}]}}

\newcommand\ul[1]{{\hat{{#1}}}}
\newcommand\covd{{\ensuremath{{\mathcal D}}}}

\newcommand\sorot{{\ensuremath{J}}}

\ifuseprd
\begin{document} % otherwise JHEP; titles, etc. in preamble
\fi %useprd

\title{A pp-Wave With 26 Supercharges}
\ifuseprd
\author{Jeremy Michelson}\email{jeremy@physics.rutgers.edu}
\altaffiliation[After Sept.\ 1, 2002: ]{Department of Physics and Astronomy,
       University of Kentucky;
       600 Rose Street;
       Lexington, KY \ 40506}\email{jeremy@pa.uky.edu}
\affiliation{New High Energy Theory Center,
       Rutgers University;
       126 Frelinghuysen Road;
       Piscataway, NJ \ 08854}
\else % use JHEP author macro
\author{
Jeremy Michelson\thanks{\tt jeremy@physics.rutgers.edu} \\
New High Energy Theory Center \\
Rutgers University \\
126 Frelinghuysen Road \\
Piscataway, NJ \ 08854 \ USA 
} 
\fi % ifuseprd for author macros

\putabstract{
A pp-wave 
solution to 11-dimensional supergravity
is given with precisely 26 supercharges.
Its uniqueness and the absence of 11-dimensional pp-waves 
which preserve (precisely) 28 or 30
supercharges is discussed.  Compactification on a spacelike circle
gives a IIA configuration with all 26 of the supercharges.  For this
compactification, D0 brane
charge does not appear in the supersymmetry algebra.
Indeed, the 26 supercharge IIA background does not admit any supersymmetric
D-branes.  In an appendix, a 28 supercharge IIB pp-wave is presented
along with its supersymmetry algebra.
}

\preprint{RUNHETC-2002-23\ifuseprd,~\else\\\fi 
   {\tt hep-th/0206204}}
%\keywords{}
%\pacs{11.25.-w, % string theory
%      04.30.-w, % grav. waves (theory)
%      04.65.+e, % SUGRA
%      04.50.+h, % gravity in >4 dimensions, Kaluza Klein, unified ft, ...
%      11.30.Pb, % supersymmetry
%     }

\ifuseprd
\maketitle
\else
\begin{document}
\fi %ifuseprd

\ifuseprd
\tableofcontents
\fi

\section{Introduction} \label{sec:intro}

Recently, supergravity solutions with the number of supercharges between 16
and 32 {\em non-inclusive\/} have been
presented.~\cite{clp,jm,clp2,ghpp,lv}  The existence of such solutions
is {\em a priori\/} surprising, although perhaps not completely
unexpected considering the results of~\cite{bj1,bj2,bjs,gh}.  It was shown
in those references that the central charge matrix appears in the
supersymmetry algebra in a way that allows for 3/4 BPS states, and it
was speculated that it should be possible to preserve any fraction,
$n/32$, of the supercharges.  However, to date the only system (to my
knowledge) in
which more than one-half---but not all---the supercharges have been
concretely observed to be
preserved is the pp-wave system; there the number of supercharges is
always even~\cite{ghpp} but has so far
been capped at 24~\cite{clp,jm,clp2,ghpp} (again not counting the maximally
supersymmetric solution~\cite{kg} or flat space).%
\footnote{After this paper was written,~\cite{br} appeared which
presents a family of
28 supercharge type IIB solutions, including the one given here
in
appendix~\ref{sec:IIB}.}
The pp-wave system is also very interesting because it provides a
class of models that are exactly solvable in string perturbation
theory~\cite{m,mt,rt}, and because of connections to Yang-Mills theory,
as first noticed in~\cite{bmn}.

It is the purpose of this
paper to present a pp-wave solution to 11-dimensional supergravity
with 26 supercharges.  This is unique and there are no 11-dimensional
pp-waves which preserve 28 or 30 supercharges.  Furthermore,
compactification can give a IIA superstring with all 26 supercharges.

A general discussion of supersymmetry in 11-dimensional supergravity
was recently given in~\cite{clp2,ghpp}.  In section~\ref{sec:setup}
we review these results, following~\cite{ghpp}, with emphasis on the
facts we will need later.  In section~\ref{sec:26susy} we present the
pp-wave which preserves 26 supercharges, and discuss the supersymmetry
algebra for the 26 supercharges.

Compactification of this background is
discussed in section~\ref{sec:compact}.  The emphasis is in
section~\ref{sec:26IIA}, on the
spacelike compactification which preserves all 26
supersymmetries---and a related compactification which preserves only
16 supersymmetries.
Compactifications and finite order orbifolds which break some of the
supersymmetry are briefly discussed
in sections~\ref{sec:SIJ} and~\ref{sec:ZNorb}, respectively.
The Green-Schwarz type IIA string with 26 supercharges,
is discussed in more detail in section~\ref{sec:IIA}.  In particular,
it is shown that the background admits no supersymmetric D-branes.

Uniqueness of the pp-wave with 26 supercharges, and a no-go
statement for 11-dimensional pp-waves with
28 and 30 supercharges, is presented in
section~\ref{sec:unique}.
In appendix~\ref{sec:IIB}, a type IIB solution with 28 supercharges is
given along with its supersymmetry algebra.

\section{The Setup} \label{sec:setup}

In ref.~\cite{clp2,ghpp}, the pp-waves of 11-dimensional supergravity
were analyzed in detail.  Here I will (partially)
follow~\cite{ghpp}. Therein is
written the ansatz
\begin{subequations} \label{pp}
\begin{align} \label{pp:ds}
ds^2 &= 2 dx^+ dx^- + \sum_{i,j} A_{ij} x^i x^j (dx^+)^2
       + \sum_i dx^i dx^i, \\
F &= dx^+ \wedge \Theta,
\end{align}
\end{subequations}
where $i=1\dots9$ and $\Theta$ is a 3-form which obeys the equation of
motion $\Tr A =
-\frac{1}{2} \norm{\Theta}^2$ and furthermore, without loss of
generality, takes either of the forms
\begin{subequations} \label{theta}
\begin{align}
\Theta %\begin{cases} 
&=
m_1 dx^1 dx^2 dx^9 + m_2 dx^3 dx^4 dx^9 
+ m_3 dx^5 dx^6 dx^9 + m_4 dx^7 dx^8 dx^9, %&
\\ \intertext{or} \begin{split} 
\Theta &= 
n_1 dx^1 dx^2 dx^3 + n_2 dx^1 dx^4 dx^5 + n_3 dx^1 dx^6 dx^7
+ n_4 dx^2 dx^4 dx^6 \\ &
\qquad \qquad + n_5 dx^2 dx^5 dx^7 + n_6 dx^3 dx^4 dx^7
+ n_7 dx^3 dx^5 dx^6.
%\end{cases}
\end{split}
\end{align}
\end{subequations}
These will be referred to as the 4-parameter or 7-parameter solutions,
respectively.  The 4-parameter solutions were analyzed in detail
in~\cite{ghpp}, and it was shown that there are solutions that
preserve 16, 18, 22, 24 or 32 supercharges.  However, the 7-parameter
solution is more difficult to analyze except in the case that, {\em
e.g.\/} $n_4=n_5=n_6=n_7=0$, for which it becomes equivalent to a
subset of the 4-parameter solution.

The analysis is conducted roughly as follows.
The Killing spinor equation is%
\begin{subequations}
\begin{equation} \label{killeq}
\covd_\mu \epsilon \equiv 
\nabla_\mu \epsilon - \Omega_\mu \epsilon = 0,
\end{equation}
where $\mu$ is a spacetime index, and
\begin{equation}
%\Omega_\mu = \frac{1}{4\cdot 144} F_{\sigma\tau\lambda\rho}
%  \left(3\Gamma^{\sigma\tau\lambda\rho}\Gamma_{\mu} - 
%	 \Gamma_\mu \Gamma^{\sigma\tau\lambda\rho}\right)\epsilon.
\Omega_\mu = \frac{1}{24} \left(3 \fs{F}\Gamma_\mu -
\Gamma_\mu \fs{F}\right); 
\quad \fs{F} \equiv \frac{1}{4!} F_{\sigma\tau\lambda\rho}
                         \Gamma^{\sigma\tau\lambda\rho}.
\end{equation}
\end{subequations}
The
$\Gamma$-matrices obey \hbox{$\anti{\Gamma^{\mu}}{\Gamma^{\nu}}=2
g^{\mu\nu}$}, and are more properly written in terms of the elfbein
\begin{align}
e^{\ul-} &= dx^{-} - \half \mu^2 [(x^I)^2+\frac{1}{4} (x^A)^2] dx^{+},&
e^{\ul+} &= dx^{+},& e^{\ul{i}}&=dx^i;
\end{align}
the hat distinguishes tangent space from spacetime indices, but
sloppiness will prevail and the hat will often be dropped.
Also, $\Gamma^{\mu\nu\cdots}$ are antisymmetrized $\Gamma$-matrices
with unit weight.
For $F = dx^+ \wedge \Theta$, it is convenient to define
\begin{equation} \label{fstheta}
\theta \equiv \fs{\Theta} \equiv \frac{1}{3!} \Theta_{ijk} \Gamma^{ijk},
\end{equation}
and
\begin{equation} \label{fstheta1}
\theta_{(i)} \equiv \Gamma^i \fs{\Theta} \Gamma^i.
\end{equation}
Note that $\theta_{(i)}$ has the same form as
$\theta$ and so $\theta$ and
$\theta_{(i)}$ all commute with each other.
With this notation,
\begin{subequations}
\begin{align}
\Omega_- &= 0, \\
\Omega_+ &= -\frac{1}{12} \fs{\Theta} (\Gamma^{+}\Gamma^{-} + \one), \\
\Omega_i &= \frac{1}{24} \Gamma^i (3 \theta_{(i)} + \theta) \Gamma^+.
\end{align}
\end{subequations}

The integrability condition for the Killing spinor
equation~\eqref{killeq} is
\begin{equation}
\com{\covd_\mu}{\covd_\nu} \epsilon = 0, \qquad \Leftrightarrow \qquad
\frac{1}{4} R_{\mu\nu\alpha\beta} \Gamma^{\alpha \beta} \epsilon
  = -\com{\Omega_\mu}{\Omega_\nu} \epsilon.
\end{equation}
Since the only non-zero components of the Riemann tensor are
\begin{equation}
R_{+i+j} = A_{ij},
\end{equation}
and since $(\Gamma^+)^2=0$, the only nontrivial components of the
integrability condition are those with $\mu=+, \nu=i$, which gives
\begin{equation} \label{intcond}
- 144 \sum_j A_{ij} \Gamma^j \Gamma^+ \epsilon
= \Gamma^i (3 \theta_{(i)}+\theta)^2 \Gamma^+ \epsilon.
\end{equation}
From equation~\eqref{intcond} it is clear that 
all pp-wave solutions of the form~\eqref{pp} preserve at least 16
supersymmetries~\cite{clp,clp2,ghpp,fp}; the corresponding Killing spinors
are annihilated by $\Gamma^+$.  Additional supersymmetries are
obtained via the following eigenvalue analysis.

The antisymmetric
matrix $\theta$ has eight (doubly degenerate%
\footnote{In~\cite{ghpp} it was convenient for the
$\Gamma$-matrices
to be
SO(9) $\Gamma$-matrices, but here it has been convenient to
use SO(1,10) $\Gamma$ matrices at the cost of doubling the degeneracy of the
eigenvalues.%
\label{ft:gamma}}%
)
skew-eigenvalues $\lambda_\Lambda$.  For example, for the 7-parameter
solution,
\begin{subequations} \label{lambda}
\begin{align}
\lambda_1 &=\phantom{-}n_1+n_2+n_3-n_4+n_5+n_6+n_7, \\
\lambda_2 &=\phantom{-}n_1+n_2+n_3+n_4-n_5-n_6-n_7, \\
\lambda_3 &=          -n_1+n_2+n_3+n_4-n_5+n_6+n_7, \\
\lambda_4 &=\phantom{-}n_1-n_2-n_3+n_4-n_5+n_6+n_7, \\
\lambda_5 &=\phantom{-}n_1-n_2+n_3+n_4+n_5-n_6+n_7, \\
\lambda_6 &=          -n_1+n_2-n_3+n_4+n_5-n_6+n_7, \\
\lambda_7 &=\phantom{-}n_1+n_2-n_3+n_4+n_5+n_6-n_7, \\
\lambda_8 &=          -n_1-n_2+n_3+n_4+n_5+n_6-n_7;
\end{align}
\end{subequations}
the skew eigenvalues for the 4-parameter solution were given in~\cite{ghpp}.
Defining the signs $s_{\Lambda a}$, $\Lambda=1\dots8,
a=1\dots7$ by
\begin{equation} \label{defs}
\lambda_\Lambda = \sum_a s_{\Lambda a} n_a,
\end{equation}
it is straightforward to see that the projection operators onto
the eigenspinors of $\theta$ with eigenvalues $\pm i \lambda_\Lambda$ are
\begin{equation} \label{proj}
\proj_\Lambda = \frac{1}{16} (\one-s_{\Lambda 1} s_{\Lambda2} \Gamma^{2345})
                       (\one-s_{\Lambda 1} s_{\Lambda3} \Gamma^{2367})
                       (\one+s_{\Lambda1} s_{\Lambda4} \Gamma^{1346})
                       (\one+s_{\Lambda1} s_{\Lambda5} \Gamma^{1357}).
\end{equation}
(There is, of course, a similar result for the 4-parameter ansatz.)

Then, to obtain additional Killing spinors, one defines the matrix
\begin{equation} \label{ui}
U_{(i)} = 3 \Gamma^i \theta \Gamma^i +\theta.
%; \theta_{(i)} = \Gamma^i \theta \Gamma^i,
\end{equation}
The eigenvalues of $U_{(i)}^2$ are doubly%
\footnote{And doubled again for SO(1,10)
$\Gamma$-matrices; see footnote~\ref{ft:gamma}.}
degenerate and given by $-\rho^2_{\Lambda(i)}$ ($\Lambda=1\dots8$).
The matrices
$U_{(i)}$ all commute with each other and with $\theta$, so the
corresponding eigenspinors
are again those which survive the projections~\eqref{proj}.
It 
follows from the integration condition~\eqref{intcond} that
if for all $i$ and some choice of $\Lambda$, we choose the matrix
$A_{ij}$ in the metric~\eqref{pp:ds} to be diagonal and given by
$A_{ij}=\mu_i^2 \delta_{ij} = \frac{1}{144} \rho_{\Lambda(i)}^2$,
then there are two more Killing spinors, namely those
not annihilated by $\Gamma^+$ but preserved by $\proj_\Lambda$.
For each $\Lambda$ for which this is satisfied, there are two additional
Killing spinors; thus the 
necessary and sufficient condition for the pp-wave to have $16+2N$ Killing
spinors is~\cite{ghpp}
\begin{equation} \label{ns}
\rho^2_{\Lambda_1(i)} = \rho^2_{\Lambda_2(i)} 
= \dots = \rho^2_{\Lambda_N(i)}, \quad \forall i=1\dots9.
\end{equation}
A necessary and useful, though insufficient, condition that follows
from this, via $i=9$, is~\cite{ghpp}
\begin{equation} \label{nis}
\lambda^2_{\Lambda_1} = \lambda^2_{\Lambda_2} 
= \dots = \lambda^2_{\Lambda_N}.
\end{equation}

The Killing spinors are given by
\begin{equation} \label{killspin}
\epsilon(\psi) = \left[\one + x^i \Omega_i\right]
                 e^{-\frac{1}{12} x^+ (\Gamma^+ \Gamma^- + \one) \theta} \psi
\end{equation}
where $\psi$ is a constant spinor which parametrizes the Killing spinor
and which obeys
\begin{equation} \label{genproj}
\half \left(\proj_{\Lambda_1} + \dots + 
\proj_{\Lambda_N}\right) \Gamma^- \Gamma^+ \psi 
+ \half \Gamma^+ \Gamma^- \psi = \psi; 
\end{equation}
that is, it survives the projection onto the subspace of the
spinors that are annihilated by $\Gamma^+$ and/or are
associated with the eigenvalues
$\lambda_{\Lambda_1},\dots,\lambda_{\Lambda_N}$.
\begin{detail}%
%\startdetail
\iftoomuchdetail
Since the $\Gamma^\pm$ matrices
in the expressions involving $e_i^*$ appear in the form of projection
operators, in principle they have tangent space indices, but of course
$\Gamma^{\ul+}=\Gamma^+$ and $\Gamma^{-}$ differs from $\Gamma^{\ul-}$
only by a piece proportional to $\Gamma^{\ul+}$ and
$(\Gamma^{\ul+})^2=0$, so $\Gamma^{-}\Gamma^+ = \Gamma^{\ul-}\Gamma^{\ul+}$.
\fi
\end{detail}%
%\finishdetail%

\section{A 26 Supercharge pp-Wave} \label{sec:26susy}

A special case of the preceding is the pp-wave
\begin{subequations} \label{26pp}
\begin{align} \label{26ppds}
ds^2 &= 2 dx^+ dx^- 
- \mu^2 \left[(x^I)^2 + \frac{1}{4} (x^A)^2\right] (dx^+)^2
+ (x^I)^2 + (x^A)^2, \begin{cases} I=1\dots 7 & \\ A = 8,9 & \end{cases} \\
\begin{split} \label{26ppf}
F &= \mu dx^+ \left[-3 dx^1 dx^2 dx^3 + dx^1 dx^4 dx^5
- dx^1 dx^6 dx^7 - dx^2 dx^4 dx^6 
\right. \\ & \quad \left.
- dx^2 dx^5 dx^7 - dx^3 dx^4 dx^7
+ dx^3 dx^5 dx^6\right].
\end{split}
\end{align}
\end{subequations}
As usual, $\mu\neq0$ can be set to any convenient value by rescaling $x^\pm$.
For this solution,
$\lambda_1=\lambda_2=-\lambda_3=\lambda_4=\lambda_8=-3\mu$, so the
condition~\eqref{nis}---and, in fact,~\eqref{ns}---is satisfied.  Thus
this solution has 26 Killing spinors, namely those preserved by the
projection operator
\begin{equation} \label{proj26}
%\half \Gamma^- \Gamma^+ 
%     \left(\proj_{1} + \proj_2 +\proj_3 + \proj_4 + \proj_{8}\right)
\proj_{12348} = 
\frac{1}{16}\left(5 \cdot \one - \frac{1}{\mu} \tilde{\theta}\right) 
      \Gamma^- \Gamma^+ 
+ \half \Gamma^+ \Gamma^-,
\end{equation}
where
\begin{equation} \label{deftt}
\tilde{\theta} = \Gamma^{1234567} \fs{\Theta}; \quad
F = dx^+ \wedge \Theta.
\end{equation}
That the projection operator~\eqref{genproj}
takes this simple form is related to equation~\eqref{nis} but still
requires some magic.
%\iftoomuchdetail
\begin{detail}%
\iftoomuchdetail
%\startdetail
(For example, it requires that
even the ``unequal'' eigenvalues are equal so that
the projection operator does in fact have eigenvalues that are only 1
or 0.  This also seems to require some sign coincidences.)
\fi
\end{detail}%
%\finishdetail%
Precisely the Killing spinors~\eqref{killspin} annihilated by
$\Gamma^+$ are independent
of $x^i$; the others depend on all of the $x^i$.

\subsection{Killing Vectors} \label{sec:killing}

The solution~\eqref{26pp} has isometry group (roughly)
$H^9\times$SU(2)$\times$SU(2)$\times$U(1), where $H$ is the Heisenberg
group.  In fact, the nine copies of the Heisenberg group share their
central element; there is also one more isometry which acts as an  outer
automorphism on the $H$.
This Heisenberg group appears for all pp-wave solutions~\cite{fp}.

For the solution~\eqref{26pp}, the
``Heisenberg'' Killing vectors are
[when there is a possibility of confusion as to whether $e_I$ is a
component or an object, we will write $k_{e_I}$ for the Killing vector]
\begin{subequations} \label{hkv}
\begin{gather}
\begin{align}
e_+ &= -\p_+, & e_- &= -\p_-, \\
e_I &= -\cos \mu x^+ \p_I - \mu x^I \sin\mu x^+ \p_-, &
e_I^* &= -\mu \sin \mu x^+ \p_I + \mu^2 x^I \cos\mu x^+ \p_-, \\
e_A &= -2\cos \frac{\mu}{2} x^+ \p_A - \mu x^A \sin\frac{\mu}{2} x^+ \p_-, &
e_A^* &= -\mu \sin \frac{\mu}{2} x^+ \p_A 
    + \frac{\mu^2}{2} x^A \cos\frac{\mu}{2} x^+ \p_-.
\end{align}
\end{gather}
\end{subequations}
The algebra is
\begin{subequations} \label{hkvalg}
\begin{gather}
\begin{align}
\com{e_I}{e^*_J} &= \mu^2 \delta_{IJ} e_-, &
\com{e_A}{e^*_B} &= \mu^2 \delta_{AB} e_-, \\
\com{e_+}{e_I} &=  e_I^*, &
\com{e_+}{e_A} &=  e_A^*, \\
\com{e_+}{e_I^*} &= -\mu^2 e_I, &
\com{e_+}{e_A^*} &= -\frac{\mu^2}{4} e_A,
\end{align}
\end{gather}
\end{subequations}
with all other commutators vanishing.  In particular,
the element $e_-$ is central.

The rotational
symmetry group of the metric is
SO(7)$\times$SO(2); however, the field strength
breaks the SO(7) to
SO(4).  Thus the symmetry group of the field configuration is
isomorphic to SU(2)$\times$SU(2)$\times$U(1).  Writing the rotational
(not necessarily Killing) vectors
\begin{equation} \label{genrot}
M_{ij} = x^i \p_j - x^j \p_i,
\end{equation}
the rotational Killing vectors are%
%\footnote{Here, $R$ stands for ``rotation'', not ``R-charge''!}
\begin{subequations}
\begin{gather}
\sorot = M_{89}, \\
\skipthis{
\begin{align}
K_1 &= M_{12}+M_{47}, & J_1 &= M_{12}-M_{56}, \\
K_2 &= M_{31}+M_{46}, & J_2 &= M_{31}-M_{75}, \\
K_3 &= M_{23}-M_{45}, & J_3 &= M_{23}+M_{67}.
\end{align}
\end{gather}
\end{subequations}
The U(1) generator is $\sorot$, and $K_x, J_x$ obey the SO(4) algebra
\begin{align}
\com{J_x}{J_y} &= \epsilon_{xyz} J_z, &
\com{K_x}{K_y} &= \epsilon_{xyz} J_z, &
\com{K_x}{J_y} &= \epsilon_{xyz} K_z.
\end{align}
The SU(2)$\times$SU(2) generators are $J_x^\pm = \half(J_x\pm K_x)$.
So,
\begin{subequations}
\begin{gather}
}% end skipthis
\begin{align}
J^+_1 &= -M_{23} + \half(M_{45}-M_{67}), &
J^-_1 &= -\half(M_{45}+M_{67}), \\
J^+_2 &= M_{31} + \half(M_{46}-M_{75}), &
J^-_2 &= -\half(M_{46}+M_{75}), \\
J^+_3 &= M_{12} + \half(M_{47}-M_{56}), &
J^-_3 &= -\half(M_{47}+M_{56}).
\end{align}
\end{gather}
\end{subequations}
The U(1) generator is $\sorot$, and $J^\pm_x$ obey the
SU(2)$\times$SU(2) algebra
\begin{align}
\com{J^+_x}{J^+_y} &= \epsilon_{xyz} J^+_z, &
\com{J^+_x}{J^-_y} &= 0, &
\com{J^-_x}{J^-_y} &= \epsilon_{xyz} J^-_z.
\end{align}

Observe that the SO(7) symmetry of the metric has decomposed as
\begin{equation} \label{decomp7}
\SO(7) \rightarrow \SO(3)\times \SO(4) \cong \SU(2)\times \SU(2) \times \SU(2)
\rightarrow \SU(2)_D\times \SU(2).
\end{equation}
In the first decomposition, SO(3) rotates $x^1,x^2,x^3$ and SO(4) rotates
$x^4,x^5,x^6,x^7$.  In the second decomposition,
$\SU(2)_D$ is the diagonal subgroup of the SO(3) and one of the
$\SU(2)\subset\SO(4)$.
While one can find that these are the rotational isometries in a
straightforward way, a nice way%
\footnote{I thank C. Hofman for this observation.}
of seeing this symmetry of the field strength
%under this
%$\SO(4)\cong \SO(3)\times \SO(3)$ 
is to notice that %the field strength can be written
\begin{equation} \label{chf}
F = \mu dx^+ \left[ -3 dx^1 dx^2 dx^3 - \sum_{y=1}^3 dx^y \wedge
\omega^-_y \right],
\end{equation}
where $\omega^-_y$ are the anti-selfdual two-forms of the $\field{R}^4$
spanned by $x^4,x^5,x^6,x^7$:
\begin{align}
\omega^-_1 &= -dx^4 dx^5 + dx^6 dx^7, &
\omega^-_2 &=  dx^4 dx^6 + dx^5 dx^7, &
\omega^-_3 &=  dx^4 dx^7 - dx^5 dx^6.
\end{align}
\skipthis{
where $\omega^+_i$ are the selfdual two-forms
\begin{align}
\omega_1^+ &= dx^{45}+dx^{67}, &
\omega_2^+ &= -dx^{46}+dx^{57}, &
\omega_3^+ &= -dx^{47}-dx^{56}.
\end{align}
}%
Note that $\lie{J_x^-}\omega^-_y = 0$ and $\lie{J_x^+}\omega^-_y =
\epsilon_{xyz}\omega^-_z$.
Equation~\eqref{chf}
is now manifestly invariant under the diagonal SO(3) (which
simultaneously rotates the $x^i$s and the $\omega^-_i$s), and under
the SO(3) that preserves the anti-selfdual two-forms.

Finally, we remark that the rotational group
\hbox{SU(2)$\times$SU(2)$\times$U(1)$\subset$SO(7)$\times$SO(2)$\subset$SO(9)}
rotates
the generators $e_i$, $e_i^*$ of the Heisenberg group in the obvious way.

\subsection{The Action of the Isometries on the Killing Spinors}

We now want to see how the isometries act on the Killing spinors.
This is useful, for example, to see what the effect of orbifolding is
on the supersymmetry.  
Recalling that the Lie derivative on spinors is given by
\begin{equation}
\lie{k} \epsilon = k^\mu \nabla_\mu \epsilon 
+ \frac{1}{4} \nabla_\mu k_\nu \Gamma^{\mu\nu} \epsilon,
\end{equation}
the reader will not be surprised to learn that
the rotations act in the usual way%
\begin{detail}%
\iftoomuchdetail
---even though a generic linear combination of $M_{ij}$s does {\em not\/}
act in this simple way!---%
\end{detail}%
\else
,
\fi
namely,
\begin{subequations} \label{jonspin}
\begin{gather}
\lie{\sorot}{\epsilon(\psi)} = \epsilon(\frac{1}{2} \Gamma^{89} \psi), \\
\begin{align}
\lie{J_1^+}{\epsilon(\psi)} &= \epsilon\left(
   \frac{1}{2}[-\Gamma^{23} + \half (\Gamma^{45}-\Gamma^{67})]\psi\right), &
\lie{J_1^-}{\epsilon(\psi)} &= \epsilon\left(
   -\frac{1}{4}[\Gamma^{45}+\Gamma^{67}]\psi\right), \\
\lie{J_2^+}{\epsilon(\psi)} &= \epsilon\left(
   \frac{1}{2}[ \Gamma^{31} + \half (\Gamma^{46}-\Gamma^{75})]\psi\right), &
\lie{J_2^-}{\epsilon(\psi)} &= \epsilon\left(
   -\frac{1}{4}[\Gamma^{46}+\Gamma^{75}]\psi\right), \\
\lie{J_3^+}{\epsilon(\psi)} &= \epsilon\left(
   \frac{1}{2}[ \Gamma^{12} + \half (\Gamma^{47}-\Gamma^{56})]\psi\right), &
\lie{J_3^-}{\epsilon(\psi)} &= \epsilon\left(
   -\frac{1}{4}[\Gamma^{47}+\Gamma^{56}]\psi\right),
\end{align}
\end{gather}
\end{subequations}
Note that because the rotations preserve $F$, therefore the linear
combinations of $\Gamma^{ij}$ that appear on the right-hand side of
equation~\eqref{jonspin}
commute with $\theta$ and the projection operator~\eqref{proj26}, and
so indeed the right-hand side is another
Killing spinor.

The Heisenberg algebra acts on the spinors as
\begin{subequations} \label{eonspin}
\begin{align}
\lie{e_-}{\epsilon(\psi)}&= 0, &
\lie{e_+}{\epsilon(\psi)}&= 
\epsilon\left(\frac{1}{12} \theta [1+\Gamma^+ \Gamma^-]\psi\right), \\
\lie{e_I}{\epsilon(\psi)}&= 
\epsilon\left(-\frac{1}{24} \Gamma^I U_{(I)}
%(3\theta \Gamma^I +\Gamma^I \theta)
      \Gamma^+ \psi \right), &
\lie{e_I^*}{\epsilon(\psi)}&= 
\epsilon\left(\frac{\mu^2}{2} \Gamma^I \Gamma^+ \psi \right), \\
\lie{e_A}{\epsilon(\psi)}&= 
\epsilon\left(-\frac{1}{12} \Gamma^A U_{(A)}
%(3\theta \Gamma^A +\Gamma^A \theta)
      \Gamma^+ \psi \right), &
\lie{e_A^*}{\epsilon(\psi)}&= 
\epsilon\left(\frac{\mu^2}{4} \Gamma^A \Gamma^+ \psi \right), &
\end{align}
\end{subequations}
where $U_{(i)}$ was defined in equation~\eqref{ui}.  Again, it is easy
to see that the algebra closes on Killing spinors%
\iftoomuchdetail
\begin{detail}
since $\Gamma^+$ annihilates all but the action of $e_+$, but $e_+$
acts on eigenspinors of $\theta$ by multiplication%
\end{detail}%
\fi
.

\begin{detail}%
%\startdetail
\iftoomuchdetail
The derivation of $\lie{e_i}{\epsilon(\psi)}$ is the following.  
Let us focus on $e_I$ for definiteness.
Since
$(\Gamma^+)^2=0$, and since $\Gamma^+$ anticommutes with $\theta$, it
is straightforward to see that
\begin{equation*}
%\begin{split}
\lie{e_I}{\epsilon(\psi)} = -\frac{1}{24} \cos\mu x^+ (3\theta
\Gamma^I + \Gamma^I \theta) e^{\frac{1}{12} x^+ \theta} \Gamma^+ \psi
- \frac{\mu}{2} \sin\mu x^+ \Gamma^I e^{\frac{1}{12} x^+ \theta}
  \Gamma^+ \psi.
%\end{split}
\end{equation*}
Thus if $\Gamma^+\psi=0$, then we are done as the right-hand side
vanishes as does the right-hand side in the expression above.
Otherwise, we must be in the subspace in which $\theta \psi = \pm 3
\mu i \psi$ (i.e.\ $\theta \Gamma^+ \psi =\mp 3 \mu i \Gamma^+ \psi$), 
for which the reader will verify, using the choice of $\mu_I$
so that supersymmetry is preserved, 
that $U_{(i)} \psi = \pm' 12 \mu i \psi$, and
therefore $\theta \Gamma^I \psi = \pm 3 \mu i \Gamma^I
\psi$, {\em or\/} $\theta \Gamma^I \psi = \mp 5 \mu i \Gamma^I \psi$,
again with the same choice of upper or lower sign.  Thus there are two
cases.  In the first case, $\theta \Gamma_I \psi = \pm 3 \mu i
\Gamma_I \psi$, $U_{(i)} \psi = \pm 12 \mu i \psi$, (i.e.\ $\pm'=\pm$) and so
\begin{equation*}
\begin{split}
\lie{e_I}{\epsilon(\psi)} &= \pm i \frac{\mu}{2} e^{\pm i \frac{9
\mu}{12} x^+} \Gamma^I \Gamma^+ \psi, \\
&= \pm i \frac{\mu}{2} e^{-\frac{1}{12} x^+ (\Gamma^+ \Gamma^- +
1)\theta} \Gamma^I \Gamma^+ \psi, \\
&= \epsilon\left(-\frac{1}{24} \Gamma^I U_{(I)} \Gamma^+ \psi\right),
\end{split}
\end{equation*}
again noting the change in sign of the eigenvalue of $U_{(I)}$ because
of the $\Gamma^+$.  For the second case, $\theta \Gamma_I \psi = \mp 5
\mu i \Gamma_I \psi$,  $U_{(i)} \psi = \mp 12 \mu i \psi$, (i.e.\
$\pm'=\mp$) and so
\begin{equation*}
\begin{split}
\lie{e_I}{\epsilon(\psi)} &= \mp i \frac{\mu}{2} e^{\pm i \frac{3\cdot5
\mu}{12} x^+} \Gamma^I \Gamma^+ \psi, \\
&= \mp i \frac{\mu}{2} e^{-\frac{1}{12} x^+ (\Gamma^+ \Gamma^- +
1)\theta} \Gamma^I \Gamma^+ \psi, \\
&= \epsilon\left(-\frac{1}{24} \Gamma^I U_{(I)} \Gamma^+ \psi\right).
\end{split}
\end{equation*}
\fi
\end{detail}%
%\finishdetail%

\subsection{The Supersymmetry Algebra}

So far we have examined the action of the bosonic symmetries.
One can check that the square of the Killing spinors is
\begin{multline} \label{ks2}
\bar{\epsilon}(\psi_1) \Gamma^\mu \epsilon(\psi_2) \p_\mu
= -(\bar{\psi}_1 \Gamma^{\ul-} \psi_2) e_- 
- (\bar{\psi}_1 \Gamma^+ \psi_2) (e_+ + \frac{\mu}{2} \sorot)
\\
%- \frac{\mu}{2} (\bar{\psi}_1 \Gamma^+ \psi_2) \sorot
- \frac{1}{24} \sum_{x=1}^3 (\bar{\psi}_1 R^{+x} \Gamma^+ \psi_2) J_x^+
+ \mu \sum_x (\bar{\psi}_1 \Gamma^{123} R^{-x} \Gamma^+ \psi_2) J_x^-
\\
- \sum_{I=1}^7 (\bar{\psi}_1 \Gamma^I \psi_2) e_I
+ \frac{1}{24 \mu^2} \sum_I [\bar{\psi}_1 (\Gamma^I U_{(I)} \Gamma^{-} \Gamma^+
                     + U_{(I)} \Gamma^I \Gamma^+ \Gamma^-) \psi_2] e_I^*
\\
- \sum_{A=8}^9 (\bar{\psi}_1 \Gamma^A \psi_2) \left(\frac{1}{2} e_A
       +\frac{1}{\mu} \sum_{B=8}^9 \epsilon_{AB} e_B^*\right)
%- \frac{1}{6 \mu^2} \sum_A (\bar{\psi}_1 \com{\Gamma^{-}}{\Gamma^{+}} \Gamma^A
%             \theta \psi_2) e_A^*,
\end{multline}
using the constant matrices
\begin{subequations} \label{mforks2}
\begin{align}
R^{+1} &= -\anti{\Gamma^{23}}{U_{(3)}}, &
R^{-1} &= \half (\Gamma^{45}+\Gamma^{67}), \\ 
R^{+2} &= \anti{\Gamma^{31}}{U_{(1)}}, &
R^{-2} &= \half (\Gamma^{46}+\Gamma^{75}), \\ 
R^{+3} &= \anti{\Gamma^{12}}{U_{(2)}}, &
R^{-3} &= \half (\Gamma^{47}+\Gamma^{56}).
\end{align}
\end{subequations}
Note that the $\Gamma$-matrices
carry tangent space indices, although this really only affects the
first term.  
\iftoomuchdetail
I%
\else
Although the derivation has been completely suppressed,%
\ \skipthis{%
\footnote{The interested reader %who wants the derivation 
can download the ``source''
of this paper from \web{http://arxiv.org/e-print/hep-th/0206204} and
uncomment line 10 of the file
(``{\tt$\backslash$toomuchdetailtrue}'') before \LaTeX{ing}.}
}%
i%
\fi
t should be noted that the sign convention
$\Gamma^{1234567} = \Gamma^{89}\Gamma^{+-}$
has been used to simplify the coefficient of $e_A^*$; see also
equation~\eqref{betterproj26} below%
\iftoomuchdetail
\begin{detail}%
\ and, in fact, the following derivation%
\end{detail}%
\fi
.

From equation~\eqref{ks2}
we see that the Killing spinors square to isometries, as
required.  Note that every isometry---and no other vector---appears on
the right-hand side.
That precisely the rotational Killing vectors appear is quite nontrivial.
However, that $e_+$ and $\sorot$ appear
only in the combination $e_+ + \frac{\mu}{2} \sorot$, and that $e_A$,
$e_A^*$ appear only in the combinations
$\half e_8 + \frac{1}{\mu} e_9^*$ and $\half e_9 -\frac{1}{\mu} e_8^*$
will have important implications for the IIA superstring; see
section~\ref{sec:26IIA}.

\iftoomuchdetail
\begin{detail}
%\startdetail
It is probably worthwhile describing how equation~\eqref{ks2} is
derived.  We start with the lengthy expression
\begin{multline}
\bar{\epsilon}(\psi_1) \Gamma^\mu \epsilon(\psi_2) \\
= \bar{\psi}_1 e^{-\frac{1}{12} x^+ (\Gamma^- \Gamma^+ + 1)\theta}
\left[\Gamma^\mu + \frac{1}{24} x^i U_{(i)} \Gamma^i \Gamma^+ \Gamma^\mu
+ \frac{1}{24} x^i \Gamma^\mu \Gamma^i U_{(i)} \Gamma^+ \right]
e^{-\frac{1}{12} x^+ (\Gamma^+ \Gamma^- + 1)\theta} \psi_2,
\end{multline}
and try to break it up into manageable chunks.  One way to break it up
is via
\begin{equation}
\psi = \half \Gamma^+ \Gamma^- \psi + \half \Gamma^- \Gamma^+ \psi
     \equiv \psi_+ + \psi_-;
\quad \Gamma^+ \psi_+ = 0 = \Gamma^{\ul-} \psi_-.
\end{equation}
Since
\begin{equation}
\bar{\psi}_\pm = \psi^\dagger \Gamma^\pm \Gamma^\mp \Gamma^0
= \pm \frac{1}{\sqrt{2}} \psi^\dagger \Gamma^\pm \Gamma^\mp 
       (\Gamma^\pm - \Gamma^\mp)
= \pm \frac{1}{\sqrt{2}} \psi^\dagger (\Gamma^\pm - \Gamma^\mp) \Gamma^\mp
       \Gamma^\pm
= \bar{\psi} \Gamma^\mp \Gamma^\pm,
\end{equation}
we also see that $\bar{\psi}_+ \Gamma^{+} = 0 = \bar{\psi}_- \Gamma^{\ul-}$.
Let us then look at the $(\pm,\pm')$ terms in turn.

For $\bar{\psi}_{1+} \dots \psi_{2+}$ we see, by inserting 
\hbox{$\one=\half \Gamma^+ \Gamma^- + \half \Gamma^- \Gamma^+$},
that the only non-zero contribution can come from $\mu={-}$, and is
therefore proportional to $-e_-$.  That is,
\begin{equation}
\bar{\epsilon}(\psi_{1+}) \Gamma^\mu \epsilon(\psi_{2+})\p_\mu
 = -\bar{\psi}_{1+} e^{-\frac{3}{4}\theta x^+} \Gamma^- 
e^{-\frac{3}{4}\theta x^+} \psi_{2+} e_-
 = -\bar{\psi}_{1+} \Gamma^- \psi_{2+} e_-,
\end{equation}
since $\theta$ anticommutes with $\Gamma^-$.  Finally, for comparison
to~\eqref{ks2}, it is obvious that $\bar{\psi}_{1}\Gamma^{\ul-}\psi_2 =
\bar{\psi}_{1+} \Gamma^{-}\psi_{2+}$.

The $e_+$ term follows similarly, but since the other $(-,-)$ terms
are rather horrible, this will be postponed until after we
are done with $(\pm,\mp)$.

Let us derive the $(+,-)$ terms in detail.
We have
\begin{equation} \label{gen+-}
\bar{\epsilon}(\psi_{1+}) \Gamma^\mu \epsilon(\psi_{2-})
= \bar{\psi}_{1+} e^{-\frac{1}{4} x^+ \theta}
\left[\Gamma^\mu 
+ \frac{1}{24} x^i \Gamma^\mu \Gamma^i U_{(i)} \Gamma^+ \right]
e^{-\frac{1}{12} x^+ \theta} \psi_{2-}.
\end{equation}
In particular, this vanishes if $\mu=+$.  The second term survives if and
only if $\mu=-$ and the first term survives if and only if $\mu=i$.
$\mu=A$ is the simplest; since $\Gamma^A$ and $\theta$ anticommute,
and since $\theta \psi_{2-}=\pm 3\mu i \psi_{2-}$---otherwise
$\psi_{2-}$ does
not parametrize a Killing spinor!---equation~\eqref{gen+-} simplifies to
\begin{equation}
\begin{split}
\bar{\epsilon}(\psi_{1+}) \Gamma^A \epsilon(\psi_{2-})
&= \bar{\psi}_{1+} \Gamma^A \psi_{2-} (\cos \frac{\mu}{2} x^+ \pm i \sin
\frac{\mu}{2} x^+) \\
&= \bar{\psi}_{1+} \Gamma^A \psi_{2-} \cos \frac{\mu}{2} x^+
  + \frac{1}{3 \mu} \bar{\psi}_{1+} \Gamma^A \theta
             \psi_{2-} \sin \frac{\mu}{2} x^+.
\end{split}
\end{equation}
For $\mu=I$, there is no simple commutation relation to move the left
exponential to the right.  Instead we note (again) that $\theta
\psi_{2-} = \pm 3 \mu i \psi_{2-}$ and $U_{(I)} \psi_{2-} = \pm' 12
\mu i \psi_{2-}$ implies either $\theta \Gamma^I \psi_{2-} =
\pm 3 \mu i \psi_{2-}$ and $\pm'=\pm$, or, $\theta \Gamma^I \psi_{2-} =
\mp 5 \mu i \psi_{2-}$ and $\pm'=\mp$.  So,
\begin{equation}
\begin{split}
\bar{\epsilon}(\psi_{1+}) \Gamma^I \epsilon(\psi_{2-})
&= \bar{\psi}_{1+} \Gamma^I \psi_{2-} (\cos \frac{\mu}{2} x^+ \mp' i \sin
\frac{\mu}{2} x^+) \\
&= \bar{\psi}_{1+} \Gamma^I \psi_{2-} \cos \frac{\mu}{2} x^+
  - \frac{1}{12 \mu} \bar{\psi}_{1+} \Gamma^I U_{(I)}
             \psi_{2-} \sin \frac{\mu}{2} x^+.
\end{split}
\end{equation}
Finally for $\mu=-$,
\begin{equation}
\begin{split}
\bar{\epsilon}(\psi_{1+}) \Gamma^- \epsilon(\psi_{2-})
&= \frac{1}{12} \sum_{i=1}^9 x^i \bar{\psi}_{1+} e^{-\frac{1}{4} x^+ \theta}
\Gamma^i U_{(i)}
e^{-\frac{1}{12} x^+ \theta} \psi_{2-} \\
&= -\frac{1}{6} \bar{\psi}_{1+} \Gamma^A \theta \psi_{2-} x^A \cos
\frac{\mu}{2} x^+
+\half \mu \bar{\psi}_{1+} \Gamma^A \psi_{2-} x^A \sin \frac{\mu}{2}
x^+
\\ & \quad
+\frac{1}{12} \bar{\psi}_{1+} \Gamma^I U_{(I)} \psi_{2-} x^I \cos
\frac{\mu}{2} x^+
+\mu \bar{\psi}_{1+} \Gamma^I \psi_{2-} x^I \sin \frac{\mu}{2} x^+.
\end{split}
\end{equation}
The analysis for the $i=A$ terms was just as before, after noting
that $U_{(A)}=-2\theta$.  For $i=I$ the analysis was also similar, so we
could immediately write the result.  Adding these up and
consulting equation~\eqref{hkv} gives
\begin{multline}
\bar{\epsilon}(\psi_{1+}) \Gamma^\mu \epsilon(\psi_{2-}) \p_\mu
= -\bar{\psi}_{1+} \Gamma^I \psi_{2-} e_I 
+ \frac{1}{12\mu^2} \bar{\psi}_{1+} \Gamma^I U_{(I)} \psi_{2-} e_I^*
\\
- \half \bar{\psi}_{1+} \Gamma^A \psi_{2-} e_A
- \frac{1}{3\mu^2} \bar{\psi}_{1+} \Gamma^A \theta \psi_{2-} e_A^*.
\end{multline}
Actually, the last term can be simplified by noting---see
equation~\eqref{betterproj26}, below---that
\begin{equation*}
\Gamma^A \theta \psi_{2-}
= -\half \epsilon_{AB} \Gamma^B \Gamma^{89} \theta \Gamma^-
\Gamma^+ \psi_{2-} = -3 \mu \epsilon_{AB} \Gamma^B \psi_{2-}.
\end{equation*}

To compute $(-,+)$, note that for Majorana spinors $\bar{\psi} =
\psi^\transpose C$, where $C=-C^\transpose$ is the charge conjugation
matrix for which $C \Gamma^\mu = -(\Gamma^\mu)^\transpose C$, 
so (for anticommuting $\psi$s)
\begin{equation}
\begin{split}
\bar{\epsilon}(\psi_{1+}) \Gamma^\mu \epsilon(\psi_{2-})
&= \bar{\psi}_{1-} e^{-\frac{1}{12} x^+ \theta}
\left[\Gamma^\mu 
+ \frac{1}{24} x^i U_{(i)} \Gamma^i  \Gamma^+ \Gamma^\mu \right]
e^{-\frac{1}{4} x^+ \theta} \psi_{2+} \\
&= -\psi_{2+}^\transpose e^{-\frac{1}{4} x^+ \theta^\transpose}
\left[\Gamma^\mu 
+ \frac{1}{24} x^i U_{(i)} \Gamma^i  \Gamma^+ \Gamma^\mu \right]^\transpose
e^{-\frac{1}{12} x^+ \theta^\transpose} C^\transpose \psi_{1-} \\
&= -\bar{\psi}_{2+} e^{-\frac{1}{4} x^+ \theta}
\left[\Gamma^\mu 
+ \frac{1}{24} x^i \Gamma^\mu \Gamma^i U_{(i)} \Gamma^+ \right]
e^{-\frac{1}{12} x^+ \theta} \psi_{1-} \\
&= -\bar{\epsilon}(\psi_{2-}) \Gamma^\mu \epsilon(\psi_{1+})
\end{split}
\end{equation}
since $\theta^\transpose C = C \theta$.  We can then change the order
again using
\begin{subequations}
\begin{align}
\bar{\psi}_{2+} \Gamma^I \psi_{1-} &= -\bar{\psi}_{1-} \Gamma^I \psi_{2+},
&
\bar{\psi}_{2+} \Gamma^A \psi_{1-} &= -\bar{\psi}_{1-} \Gamma^A \psi_{2+},
\\
\bar{\psi}_{2+} \Gamma^I U_{(I)} \psi_{1-} &= 
-\bar{\psi}_{1-} U_{(I)} \Gamma^I \psi_{2+}, &
\bar{\psi}_{2+} \Gamma^A \theta \psi_{1-} &= 
\bar{\psi}_{1-} \Gamma^A \theta \psi_{2+}.
\end{align}
\end{subequations}
Now we can combine the $(-,+)$ and $(+,-)$ terms; getting signs
(especially for the $e_A^*$ terms) and ordering (for the $e_I^*$
terms) right accounts for the projection operators---or lack
thereof---in equation~\eqref{ks2}.  Note that although the derivation
used anticommuting spinors, because we have reverted to the original
ordering, the result~\eqref{ks2} holds for commuting spinors as well.

Finally, we can no longer postpone the $(-,-)$ terms.  These read
\begin{equation}
\begin{split}
\bar{\epsilon}(\psi_{1-}) \Gamma^\mu \epsilon(\psi_{2-}) \p_\mu
&= \bar{\psi}_{1-} e^{-\frac{1}{12} x^+ \theta}
\left[\Gamma^\mu 
+ \frac{1}{24} x^i \Gamma^\mu  \Gamma^i U_{(i)} \Gamma^+
+ \frac{1}{24} x^i U_{(i)} \Gamma^i  \Gamma^+ \Gamma^\mu \right]
e^{-\frac{1}{12} x^+ \theta} \psi_{2-} \p_\mu \\
&= \bar{\psi}_{1-} e^{-\frac{1}{12} x^+ \theta}
\left[\Gamma^\mu 
+ \frac{1}{24} x^i \anti{\Gamma^{\mu i}}{U_{(i)}} \Gamma^+
\right]
e^{-\frac{1}{12} x^+ \theta} \psi_{2-} \p_\mu,
\end{split}
\end{equation}
where in the second step, it was noted that the $\mu=-$ contribution
vanishes, and that all but the first term vanishes if $\mu=+$.  In
fact, as for the $(+,+)$ terms, the first term contributes if and only
if $\mu=+$, for which we obtain
\begin{equation}
\bar{\epsilon}(\psi_{1-}) \Gamma^+ \epsilon(\psi_{2-}) \p_+
= -\bar{\psi}_{1-} \Gamma^+ \psi_{2-} e_+.
\end{equation}
The remaining terms involve $x^i \Gamma^{ji} \p_j$ and will therefore
be proportional to $M_{ij}$.  However, the $M_{ij}$s are not Killing
vectors (except for $M_{89}$) and so we must figure out how the terms
combine into precisely the SO(4)$\times$SO(2) Killing vectors.
Let us
start by considering $\mu=A, i=B$.  This is easy,
\begin{equation}
\begin{split}
\frac{1}{24} x^B \bar{\psi}_{1-} e^{-\frac{1}{12} x^+ \theta}
\anti{\Gamma^{AB}}{U_{(B)}} \Gamma^+
e^{-\frac{1}{12} x^+ \theta} \psi_{2-} \p_A
&= -\frac{1}{6} \epsilon^{AB} x^B 
\bar{\psi}_{1-} \Gamma^{89} \theta \Gamma^+ \psi_{2-} \p_A \\
&= \frac{\mu}{2} \epsilon^{AB} x^B \bar{\psi}_{1-} \Gamma^+ \psi_{2-} \p_A,
\end{split}
\end{equation}
with no sum over $A,B$.
In the last step we have anticommuted $\Gamma^{89} \theta$
through $\Gamma^+$ and then taken advantage of the
projection~\eqref{proj26}, in the form~\eqref{betterproj26} below.
Summing then produces $-\frac{\mu}{2} \bar{\psi}_{1-} \Gamma^+
\psi_{2-} M_{89}$.

A contribution from $\mu=A, i=I$ is forbidden.  Indeed, note
\begin{equation*}
\anti{\Gamma^{AI}}{U_{(I)}} = \anti{\Gamma^A}{U_{(I)}} \Gamma^I 
+ \Gamma^A \com{\Gamma^I}{U_{(I)}} = 2\Gamma^A\com{\theta}{\Gamma^I}.
\end{equation*}
Again we use the fact that, with $\theta \psi_- = \pm 3 \mu i \psi_-$,
then $\theta \Gamma^I \psi = \begin{cases} \pm 3 \mu i \Gamma^I \psi_-,
\\ \mp 5 \mu i \Gamma^I \psi_-. \end{cases}$  The first case implies
$\com{\theta}{\Gamma^I} \psi_- = 0$.  The second case implies that the
left-hand exponential is $e^{\mp \frac{5}{12} \mu i x^+}$ (with an
extra sign from $\Gamma^+$); on the
other hand, if we had used the eigenvalue of $\psi_{1-}$, we would have
obtained $e^{\mp' \frac{3}{12}\mu i x^+}$ so consistency requires that
the expression vanish.  This standard
argument will be used several times below.

This leaves $\mu=I$, $i=J$.
As a
temporary aside, note that
\begin{equation} \label{gijuj}
\anti{\Gamma^{IJ}}{U_{(J)}} = 3 \Gamma^I \theta \Gamma^J 
- 3 \Gamma^J \theta \Gamma^I + \Gamma^{IJ} \theta
+ \theta \Gamma^{IJ} = \anti{\Gamma^{IJ}}{U_{(I)}}.
\end{equation}
We will evaluate a few cases and
argue that symmetry implies the general result.
First we will argue that all the terms with
$\mu=I=1\dots3$ and $i=J=4\dots7$ (or vice versa) vanish; this is
required by the absence of $M_{1\dots3,4\dots7}$ in the isometry
group.  It will suffice to consider $I=1,J=4$; the other cases are
obtained by utilizing the SO(4) symmetry.  That is, once we know that
$I=1,J=4$ does not contribute, then we can use $J^-_x$ to conclude
that none of $I=1,J=4\dots7$ contribute, from which $J^+_x$ implies
that none of $M_{1\dots3,4\dots7}$ appear.
Straightforward computation gives
\begin{equation} \label{antifor14}
\anti{\Gamma^{14}}{U_{(1)}} = -4\mu \Gamma^5
\left[(\one-\Gamma^{1247}) + (\one+\Gamma^{1346})\right].
\end{equation}
Comparing to the projection operator~\eqref{proj}---or the
eigenvalues~\eqref{lambda}; e.g.\ $\Gamma^{1247} =
\Gamma^{123}\Gamma^{347}$ implies that $\Gamma^{1247} \proj_1 = -\proj_1$
(since $s_{11} s_{16}=1$ and $i^2=-1$), $\Gamma^{1247}\proj_2=\proj_2$, 
etc.---we see that these
project out the spinors preserved by $\proj_3$; the quantity in square
brackets is just a number for the other (basis) spinors.  However, further
straightforward computation, using~\eqref{lambda}, determines that it is
precisely the remaining spinors (that is those preserved by
$\proj_1+\proj_2+\proj_4+\proj_8$; the other 6 are irrelevant is they
do not parametrize Killing spinors) for which $\theta \Gamma^5 \psi_-
= \mp 5 i \mu \Gamma^5 \psi_-$.  This gives us our promised
contradiction when we evaluate the left-hand exponential by either
working to the left [so acting $\theta$ on $\Gamma^5 \psi$, using
equation~\eqref{antifor14}] or working to the right.  So the full
matrix element vanishes.  The same computation shows that $\mu=4,i=1$
also does not
contribute.

It remains to show that those $\mu=I,i=J$ terms which do contribute,
do so in a way so that the $M_{IJ}$ sum into $J^\pm_x$.  This time we
will focus on $\mu=1, i=2$, $\mu=4, i=7$ and $\mu=5,i=6$, and the
similar [via equation~\eqref{gijuj}]
terms from $\mu\rightarrow i$; these should
combine into $J^\pm_3$.  The remaining $J^\pm_x$ then follow from the
SO(4) symmetry.
Again, we resort to direct computation and note that
\begin{subequations} \label{anticomsforj3}
\begin{align}
\anti{\Gamma^{12}}{U_{(2)}} &= 16\mu \Gamma^3 + 
4 \mu \Gamma^3 \left[(\one + \Gamma^{1247}) +
(\one-\Gamma^{1256})\right], \\
\anti{\Gamma^{47}}{U_{(7)}} &= 
4 \mu \Gamma^3 \left[3(\one + \Gamma^{1247}) -
(\one-\Gamma^{4567})\right], \\
\anti{\Gamma^{56}}{U_{(6)}} &= 
-4 \mu \Gamma^3 \left[3(\one - \Gamma^{1256}) +
(\one-\Gamma^{4567})\right].
\end{align}
\end{subequations}
Comparing to the projection operators~\eqref{proj}---or better yet,
the original eigenvalues~\eqref{lambda}---shows us that
\begin{subequations}
\begin{align}
\anti{\Gamma^{12}}{U_{(2)}} \proj_1 &= 24 \mu \Gamma^3 \proj_1, &
\anti{\Gamma^{56}}{U_{(6)}} \proj_1 &= -24 \mu \Gamma^3 \proj_1, &
\anti{\Gamma^{47}}{U_{(7)}} \proj_1 &= 0, \\
\anti{\Gamma^{12}}{U_{(2)}} \proj_2 &= 24 \mu \Gamma^3 \proj_2, &
\anti{\Gamma^{56}}{U_{(6)}} \proj_2 &= 0, &
\anti{\Gamma^{47}}{U_{(7)}} \proj_2 &= 24 \mu \Gamma^3 \proj_2, \\
\anti{\Gamma^{12}}{U_{(2)}} \proj_3 &= 24 \mu \Gamma^3 \proj_3, &
\anti{\Gamma^{56}}{U_{(6)}} \proj_3 &= 0, &
\anti{\Gamma^{47}}{U_{(7)}} \proj_3 &= 24 \mu \Gamma^3 \proj_3, \\
\anti{\Gamma^{12}}{U_{(2)}} \proj_4 &= 24 \mu \Gamma^3 \proj_4, &
\anti{\Gamma^{56}}{U_{(6)}} \proj_4 &= -24 \mu \Gamma^3 \proj_4, &
\anti{\Gamma^{47}}{U_{(7)}} \proj_4 &= 0, \\
\anti{\Gamma^{12}}{U_{(2)}} \proj_8 &= 32 \mu \Gamma^3 \proj_8, &
\anti{\Gamma^{56}}{U_{(6)}} \proj_8 &= -16 \mu \Gamma^3 \proj_8; &
\anti{\Gamma^{47}}{U_{(7)}} \proj_8 &= 16 \mu \Gamma^3 \proj_8, 
\end{align}
\end{subequations}
we thus see that
\begin{equation} \label{anticomrelated}
\anti{\Gamma^{47}}{U_{(7)}} \psi_- = \anti{\Gamma^{12}}{U_{(2)}} \psi_-
+ \anti{\Gamma^{56}}{U_{(6)}} \psi_-,
\end{equation}
and so, on Killing spinors not annihilated by $\Gamma^+$,
\begin{equation}
\begin{split}
\anti{\Gamma^{12}}{U_{(2)}} M_{12} &+ \anti{\Gamma^{47}}{U_{(7)}} M_{47}
+ \anti{\Gamma^{56}}{U_{(6)}} M_{56} \\
&= \anti{\Gamma^{12}}{U_{(2)}} (M_{12}+M_{47})
 + \anti{\Gamma^{56}}{U_{(6)}}(M_{56}+M_{47}) \\
&= \anti{\Gamma^{12}}{U_{(2)}} J^+_3 
- \left( 2 \anti{\Gamma^{56}}{U_{(6)}}+ \anti{\Gamma^{12}}{U_{(2)}}\right)
   J^-_3 \\
&= \anti{\Gamma^{12}}{U_{(2)}} J^+_3 
- \left( \anti{\Gamma^{47}}{U_{(7)}} + \anti{\Gamma^{56}}{U_{(6)}} \right)
   J^-_3 \\
&= \anti{\Gamma^{12}}{U_{(2)}} J^+_3 
- 12 \mu \Gamma^{123} (\Gamma^{47}+\Gamma^{56}) J^-_3,
\end{split}
\end{equation}
where we have used equations~\eqref{anticomrelated}
and~\eqref{anticomsforj3}, as well as the obvious $M_{47} =
\half(M_{47}+M_{56})+\half(M_{47}-M_{56})$ which was used to get from
the second line to the third line.  Again, we emphasize that this
holds only on the 10 spinors preserved by $\half \proj_{12348}
\Gamma^- \Gamma^+$.  But that is enough to plug in and find
\begin{equation}
\bar{\epsilon}(\psi_{1-}) \Gamma^I \epsilon(\psi_{2-}) \p_I
= -\frac{1}{24} \bar{\psi_{1-}} R^{+x} \Gamma^+ \psi_{2-} J^+_x
  + \mu \bar{\psi_{1-}} \Gamma^{123} R^{-x} \Gamma^+ \psi_{2-}
        J^-_x.
\end{equation}
This finally completes the derivation of equation~\eqref{ks2}.

\fi
\end{detail}% note paragraph break after the detail
%\finishdetail%
% already did a paragraph break before the detail!
\skipthis{% test detail environment with (sub)equations
\begin{subequations}
\begin{align}
1=2-1
\end{align}
%\begin{detail}%
\startdetail
\begin{align}
3=4-1
\end{align}
%\end{detail}%
\finishdetail%
\begin{align}
17=16+1
\end{align}
\end{subequations}
}% end skipthis
The full supersymmetry algebra can now be written.
%\begin{subequations}
\begin{gather} \label{QQ}
\begin{split}
\anti{Q_\alpha}{Q_\beta} =& 
-(\Gamma^{-} C^{-1})_{\alpha\beta} e_-
-(\Gamma^{+} \proj_{12348} C^{-1})_{\alpha\beta} (e_+ + \frac{\mu}{2} \sorot)
%-\frac{\mu}{2} (\Gamma^+ C^{-1})_{\alpha\beta} \sorot
\\ &
- \frac{1}{24} \sum_{x=1}^3 (R^{+x} \Gamma^+ \proj_{12348} 
               C^{-1})_{\alpha\beta} J^+_x
+ \mu \sum_{x=1}^3 (\Gamma^{123} R^{-x} \Gamma^+ C^{-1})_{\alpha\beta} J^-_x
\\ &
- \sum_{I=1}^7 (\Gamma^I \proj_{12348} C^{-1})_{\alpha\beta} e_I
+ \frac{1}{24\mu^2} \sum_{I=1}^7 \left[(\Gamma^I U_{(I)} \Gamma^- \Gamma^+
            + U_{(I)} \Gamma^I \Gamma^+ \Gamma^-)C^{-1}\right]_{\alpha\beta}
       e^*_I
\\ &
- \sum_{A=8}^9 (\Gamma^A \proj_{12348} C^{-1})_{\alpha\beta} \left(\half e_A +
  \frac{1}{\mu} \sum_B \epsilon_{AB} e_B^*\right),
%- \frac{1}{6 \mu^2} \sum_{A=8}^9 \left[(\Gamma^-\Gamma^+ - \Gamma^+\Gamma^-)
%       \Gamma^A \theta C^{-1}\right]_{\alpha\beta} e_A^*,
\end{split}% \\
\end{gather}
\begin{subequations}
\begin{gather}
\begin{align}
\com{e_-}{Q} &= 0, &
\com{e_+}{Q} &= \frac{1}{12} \theta (\one + \Gamma^- \Gamma^+) Q, \\
\com{e_I}{Q} &= -\frac{1}{24} \tilde{\proj}_{12348} U_{(I)} 
                       \Gamma^I \Gamma^+ Q, &
\com{e_I^*}{Q} &= -\frac{\mu^2}{2} \tilde{\proj}_{12348} 
                       \Gamma^I \Gamma^+ Q, \\
\com{e_A}{Q} &= -\frac{1}{12} \tilde{\proj}_{12348} U_{(A)} 
                       \Gamma^A \Gamma^+ Q, &
\com{e_A^*}{Q} &= -\frac{\mu^2}{4} \tilde{\proj}_{12348} 
                       \Gamma^A \Gamma^+ Q, \\
\com{J^+_1}{Q} &= -\half \left[-\Gamma^{23} 
            +\half (\Gamma^{45}-\Gamma^{67})\right] Q, &
\com{J^-_1}{Q} &= \frac{1}{4} (\Gamma^{45}+\Gamma^{67}) Q, \\
\com{J^+_2}{Q} &= -\half \left[ \Gamma^{31} 
            +\half (\Gamma^{46}-\Gamma^{75})\right] Q, &
\com{J^-_2}{Q} &= \frac{1}{4} (\Gamma^{46}+\Gamma^{75}) Q, \\
\com{J^+_3}{Q} &= -\half \left[ \Gamma^{12} 
            +\half (\Gamma^{47}-\Gamma^{56})\right] Q, &
\com{J^-_3}{Q} &= \frac{1}{4} (\Gamma^{47}+\Gamma^{56}) Q,
\end{align} \\
\com{\sorot}{Q} = -\half \Gamma^{89} Q,
\end{gather}
\end{subequations}%
\begin{subequations}
\begin{gather} \label{bosoniccoms}
\begin{align}
\com{e_I}{e^*_J} &= \mu^2 \delta_{IJ} e_-, &
\com{e_A}{e^*_B} &= \mu^2 \delta_{AB} e_-, \\
\com{e_+}{e_I} &= e_I^*, &
\com{e_+}{e_A} &= e_A^*, \\
\com{e_+}{e_I^*} &= -\mu^2 e_I, &
\com{e_+}{e_A^*} &= -\frac{\mu^2}{4} e_A, \\
\com{J^+_x}{J^+_y} &= \epsilon_{xyz} J^+_z, &
\com{J^-_x}{J^-_y} &= \epsilon_{xyz} J^-_z. \\
\com{J^+_x}{e_I} &= \sum_{J=1}^7 R^{+x}_{IJ} e_J, &
\com{J^+_x}{e^*_I} &= \sum_{J=1}^7 R^{+x}_{IJ} e^*_J, \\
\com{J^-_x}{e_I} &= \sum_{J=1}^7 R^{-x}_{IJ} e_J, &
\com{J^-_x}{e^*_I} &= \sum_{J=1}^7 R^{-x}_{IJ} e^*_J, \\
\com{\sorot}{e_A} &= \delta_{A,9} e_8 - \delta_{A,8} e_9, &
\com{\sorot}{e^*_A} &= \delta_{A,9} e^*_8 - \delta_{A,8} e^*_9,
\end{align}
\end{gather}
\end{subequations}%
where, in terms of the rotation matrices $M^{IJ}_{KL} = \delta^I_K
\delta^J_L - \delta^I_L \delta^J_K$, the matrices
in~\eqref{bosoniccoms} are
\begin{subequations}
\begin{align}
R^{+1}_{IJ} &= -M^{23}_{IJ}+\half ( M^{45}_{IJ}-M^{67}_{IJ}), &
R^{-1}_{IJ} &= -\half ( M^{45}_{IJ}+M^{67}_{IJ}), \\
R^{+2}_{IJ} &=  M^{31}_{IJ}+\half ( M^{46}_{IJ}-M^{75}_{IJ}), &
R^{-2}_{IJ} &= -\half ( M^{46}_{IJ}+M^{75}_{IJ}), \\
R^{+3}_{IJ} &=  M^{12}_{IJ}+\half ( M^{47}_{IJ}-M^{56}_{IJ}), &
R^{-3}_{IJ} &= -\half ( M^{47}_{IJ}+M^{56}_{IJ}).
\end{align}
\end{subequations}%
All other commutators vanish.
Here, $\alpha,\beta,\dots$ are spinor indices; $C$ is the charge
conjugation matrix; and $\tilde{\proj}_{12348}$ is the projection operator
\begin{equation}
\tilde{\proj}_{12348} = C^{-1} \proj_{12348}^\transpose C
= \frac{1}{16}\left(5 \cdot \one - \frac{1}{\mu} \tilde{\theta}\right) 
      \Gamma^+ \Gamma^-
+ \half \Gamma^- \Gamma^+.
\end{equation}
The same symbols have been used here for the bosonic
charges as were used for their Killing vectors, and $Q$ are the supercharges%
\iftoomuchdetail
\begin{detail}
defined so that, say, the
supersymmetry transformation of the gravitino $\Psi_\mu$ is
$\delta \Psi_\mu = \com{\bar{\psi}_\alpha Q_\alpha}{\Psi_\mu} = 
\covd_\mu \epsilon(\psi)$, where $\covd_\mu$ is the supercovariant
derivative~\eqref{killeq}%
\end{detail}%
\fi%
; note that $\anti{\bar{\psi}_1 Q}{\bar{\psi}_2 Q} 
= \bar{\epsilon}(\psi_1) \Gamma^\mu \epsilon(\psi_2) \p_\mu$,
and \hbox{$\com{B}{\bar{\psi}Q} = \bar{\psi} C^{-1} M^\transpose C Q 
= \bar{\psi} \tilde{\proj}_{12348} C^{-1} M^\transpose C Q$} if
$\lie{B} \epsilon(\psi) = \epsilon(M\psi)$.

\section{Orbifolds} \label{sec:compact}

As in~\cite{jm}, we can use the isometries to find supersymmetric
spatial compactifications of the pp-wave.  Explicitly, define
%\begin{subequations}
\begin{align}
S_{AB}^\pm &= \frac{1}{2}e_A \pm \frac{1}{\mu} e_B^*, &
S_{IJ}^\pm &= e_I \pm \frac{1}{\mu} e_J^*, 
%\\
%S_{AI}^\pm &= \frac{1}{2}e_A + \frac{1}{\mu} e_I^*, &
%S_{IA}^\pm &= e_I + \frac{1}{\mu} e_A^*,
\end{align}
%\end{subequations}
and note that $\norm{S_{IJ}^\pm}^2 = 1 = \norm{S_{AB}}^2$ 
($I\neq J, A\neq B$).  Of course one could
compactify on any linear combination of the isometries; however, given
the results of~\cite{jm},
these are the isometries for which one expects a chance of respecting
some supersymmetry above the 16 annihilated by $\Gamma^+$.
%As we will see, however, this expectation requires some modification.

\subsection{A Compactification with 26 Supercharges} \label{sec:26IIA}

The simplest compactification is along $S^\pm_{AB} (A\neq B)$; this is
the simplest because $\theta$ is free of $\Gamma^A$.  From this fact and
equation~\eqref{eonspin},
\begin{equation}
\lie{S^\pm_{AB}}{\epsilon(\psi)}
= \epsilon(\frac{1}{12} \Gamma^A (\theta\mp 3 \mu \Gamma^{AB}) \Gamma^+
\psi).
\end{equation}
So if 
\begin{equation}
\Gamma^+ \psi \stackrel{?}{=} 0 \quad \text{or} \quad 
(\theta\Gamma^{89} \pm 3 \mu \epsilon_{AB} )\psi \stackrel{?}{=} 0, 
\end{equation}
then $\psi$ parametrizes a supersymmetry that is
preserved by compactification on $S^\pm_{AB}$, provided $\psi$ is also
in the 26-dimensional subspace of constant spinors preserved by the projection
operator~\eqref{proj26}.  That projection keeps all the $\Gamma^+
\psi=0$ spinors, and also the
\begin{equation} \label{betterproj26}
(5\cdot \one - \frac{1}{\mu} \tilde{\theta})\Gamma^- \Gamma^+ \psi = 
\psi \quad \Leftrightarrow 
(3 - \frac{1}{\mu} \theta \Gamma^{89}) \Gamma^- \Gamma^+ \psi = 0,
\end{equation}
using $\Gamma^{1234567}=\Gamma^{89}\Gamma^{+-}$.  Thus we see that %\\
\begin{center}
\begin{tabular}{rl}
$S^+_{89}, S^-_{98}$ & preserve the 16 supersymmetries annihilated by
$\Gamma^+$, \\
$S^-_{89}, S^+_{98}$ & preserve all 26 supersymmetries.
\end{tabular}
\end{center}
It might seem strange that $S^+_{AB}$ preserves a different number of
supersymmetries from $S^-_{AB}$---na\"{\i}vely they are equivalent via
parity ($x^8\rightarrow -x^8$, say)---so it should be emphasized that
which of $S^+_{AB}$ or $S^-_{AB}$ is the fully supersymmetric circle
depends on whether we choose \hbox{$\Gamma^{1234567} = \pm \Gamma^{89}
\Gamma^{+-}$}.  This partial chirality of the solution is rather curious.

The Type IIA configuration that preserves 26 supercharges is
\begin{subequations} \label{26IIA}
\begin{align}
ds^2 &= 2 dX^+ dX^- - \mu^2 (X^i)^2 (dX^+)^2 + (dX^i)^2, \quad i=1\dots8,\\
^{(2)}F &= -\mu dX^+ dX^8, \\
^{(4)}F &= -\mu dX^+ \left[ 3 dX^1 dX^2 dX^3 + dX^y \wedge
\omega_y^-\right],
\end{align}
\end{subequations}
where $^{(2)}F$ is the Kaluza-Klein, \RR, field strength.
This is derived as follows.  From equation~\eqref{26pp}, make the
change of coordinates
\begin{subequations} \label{xtoX}
\begin{gather}
\begin{align}
x^+ &= X^+, & x^- &= X^- + \frac{\mu}{2} X^8 X^9, & x^I &= X^I,
\end{align} \\ \label{x8toX8}
\begin{align} 
x^8 &= X^9 \cos (\frac{\mu}{2} X^+) + X^8 \sin (\frac{\mu}{2} X^+), &
x^9 &= -X^9 \sin (\frac{\mu}{2} X^+) + X^8 \cos (\frac{\mu}{2} X^+).
\end{align}
\end{gather}
\end{subequations}
\iftoomuchdetail%
\begin{detail}%
Note that this coordinate transformation, restricted to the transverse
coordinates, is an $X^+$-dependent {\em improper\/} rotation.
\end{detail}%
\fi%
Then $\frac{\p}{\p X^9} = -S^-_{89}$ is a manifest Killing direction;
the metric reads
\begin{equation} \label{s-89}
ds^2 = 2 dX^+ dX^- - \mu^2 \sum_{i=1}^8 (X^i)^2 (dX^+)^2
         + \sum_{i=1}^8 (dX^i)^2 + (dX^9 + \mu X^8 dX^+)^2.
\end{equation}
This and standard dimensional reduction formul\ae---namely
\hbox{$ds_{11}^2 = ds^2_{10} + (dx^9 + A_\mu dx^\mu)^2$}---then give the
configuration~\eqref{26IIA}.

The Killing spinors
are independent of the internal $X^9$
coordinate.  Thus the perturbative type IIA string in this background
sees 26 supercharges.  However, note that the Killing vector which
generates the circle $S^-_{89}$ (or $S^+_{98}$) does {\em not\/}
appear on the right-hand side of the supersymmetry
algebra~\eqref{QQ}.  Thus, this IIA string does not admit
supersymmetric D0 branes.  Indeed if this 26 supercharge IIA string
did admit supersymmetric D0 branes---that is, if $S^-_{89}$ and
$S^+_{98}$ appeared in the square of the supercharges---then there
would be a
contradiction with the Jacobi identity since $\com{S^-_{89}}{S^+_{98}}\neq 0$.
Similarly, observe that the combination $-e_+ - \frac{\mu}{2} \sorot$
which appears in the algebra~\eqref{QQ} is just the Hamiltonian,
$\frac{\p}{\p X^+}$, of the new coordinates~\eqref{xtoX}, and the
$\SO(2)$ generator $\sorot$, that is broken by the
compactification, does not appear in the 10-dimensional supersymmetry algebra.

Finally, note that we cannot compactify on another (commuting and
spacelike) circle
without breaking some of the supersymmetry (see section~\ref{sec:SIJ}).

\subsection{Other Compactifications} \label{sec:SIJ}

If we compactify on any of the individual isometries $S^{\pm}_{IJ}$, 
at least the 16 supercharges annihilated by $\Gamma^+$ are preserved.  
\skipthis{
However,
if we consider more general linear combinations of these isometries,
then more supersymmetries may be preserved.
}%
Unfortunately, the
general situation is rather difficult to analyze.  It appears,
however, that some of the 26 supercharges are always broken.

Specifically, note that
\begin{equation}
\lie{S^\pm_{IJ}} \epsilon(\psi) 
= \epsilon\left(-\frac{1}{24} \Gamma^I (U_{(I)} \mp 12 \mu \Gamma^{IJ})
        \Gamma^+ \psi \right).
\end{equation}
Unlike the isometries discussed in section~\ref{sec:26IIA}, there is
no ``chirality'' condition on the spinors that allows this to vanish
for all 26 supersymmetries, even upon taking linear combinations.
The
simplest way to understand this is to note that every circle will break
at least some of the $\SO(4)$ rotational symmetries.  Since these appear on
the right-hand side of~\eqref{QQ}, the Jacobi identity requires
that some of the supersymmetry also be broken by the circle.
Generically, therefore, only 16 supersymmetries are preserved.
However, note, for example, that a circle along
\begin{center}
\begin{tabular}{rl}
$S^\pm_{12}$ & preserves 20 supercharges, \\
$S^+_{45} + S^-_{67}$ & preserves 20 supercharges, \\
$S^+_{45} - S^+_{67}$ & preserves 20 supercharges, \\
$S^\pm_{45}$ & preserves 18 supercharges, \\
$S^+_{45} - S^-_{67}$ & preserves 18 supercharges, \\
$S^+_{12} + S^-_{45}+S^-_{67}$ & preserves 18 supercharges,\\
$S^-_{45} + S^-_{67}$ & preserves 16 supercharges,\\
$S^+_{12} + S^-_{45}-S^\pm_{67}$ & preserves 16 supercharges.
\end{tabular}
\end{center}

\subsection{Finite Orbifolds} \label{sec:ZNorb}

We can also find supersymmetric orbifolds by considering finite
subgroups \hbox{$\Gamma\subset \SU(2)\times\SU(2)$}.  In flat space---as well
as for pp-waves with more conventional (16 or 32) numbers of
supersymmetries~\cite{rt}---$\Gamma\subset \SU(2)_-$ preserves half of
the supercharges.
Therefore, it will preserve at least eight supercharges---that is half
of the sixteen conventional supercharges annihilated by $\Gamma^+$.
For simplicity, we will only consider $\ZZ_N$ orbifolds here; then
direct computation shows that $\ZZ_N\subset \SU(2)_-$ generically
preserves the 10
supercharges for which
\begin{equation}
\proj_{12348} \psi = \psi, \quad \text{but}, \quad
(\Gamma^{45}+\Gamma^{67}) \psi  = 0,
\end{equation}
having chosen the $\ZZ_N$ to be along the $J^-_1$ orbit.  Note that
the second condition picks out 16 of the 32 SO(10,1) spinors, but that
the first condition acts asymmetrically on these, and keeps 10 of
them.

For $\ZZ_N\subset \SU(2)_+$, we should realize that each $J^+_x$ acts
simultaneously on three two-planes.  Therefore, we are only guaranteed 4
supercharges.
In fact, the $\ZZ_N$ generated by $J^+_1$ generically
preserves 6 supercharges.

Finally, the $\ZZ_N\subset SU(2)_D \subset SU(2)_+\times
SU(2)_-$, generated by 
\hbox{$J^+_1-J^-_1 = -M_{23}+M_{45}$}, generically preserves 14
supercharges, which is 6 more than the guaranteed eight.

\section{The 26 Supercharge IIA Background} \label{sec:IIA}

Equation~\eqref{26IIA} gives a type IIA background with 26
supersymmetries.  Dimensionally reducing the Killing spinor
equation~\eqref{killeq}---and, in the $X^\mu$ coordinate system,%
\footnote{One should beware that 
because equation~\eqref{x8toX8} is an improper rotation,
there is a sign change in the $X^\mu$ coordinate system to
$\Gamma^{1234567} = -\Gamma^{89} \Gamma^{+-}$ as compared to the
discussion surrounding equation~\eqref{betterproj26}.  As a result,
$\Gamma^{11} = \Gamma^{+-12345678} = +\Gamma^9$, in these
conventions, yet equation~\eqref{betterproj26} reads
\hbox{$(3+\frac{1}{\mu}
\fs{\Theta}\Gamma^8 \Gamma^{11})\Gamma^{-}\Gamma^+ \psi=0$}.}
setting the chirality matrix
$\Gamma^{11}= \Gamma^9$---gives the
IIA Killing spinor equation,
\begin{equation} \label{scdIIA}
{\mathcal D}_\mu \epsilon = \nabla_\mu \epsilon
+ \frac{1}{4} {^{(2)}F}_{\mu\nu} \Gamma^\nu \Gamma^{11} \epsilon
 - \frac{1}{24} (3 \fs{^{(4)}F} \Gamma_\mu 
          - \Gamma_\mu \fs{^{(4)}F} ) \epsilon
\equiv \nabla_\mu \epsilon - \Omega_\mu \epsilon.
\end{equation}
The integrability of the 26 Killing spinors is both guaranteed, and
easy to check.

It is now straightforward, using the shortcut of~\cite{mt}---which can
be checked against the
results of~\cite{clps}---to write down the lightcone gauge-fixed
Green-Schwarz action in this background.  It is convenient to set
$\apr=1=p^+$.
Since the two Majorana-Weyl spinors, $S^\Lambda$, of the Green-Schwarz
string are of opposite
chirality, we combine them into a single Majorana
spinor $S$.  
\skipthis{
The two chiralities are easily extracted if we use a Majorana-Weyl
basis for the $\Gamma$-matrices for which $\Gamma^{+,-,I} =
\gamma^{+,-,I}\otimes \sigma^2$, $\Gamma^8 = \one \otimes \sigma^1$,
$\Gamma^{11} = \one \otimes \sigma^3$, using the
Majorana (pure imaginary) $\gamma$-matrices of $\SO(7,1)$.}%
Lightcone gauge is defined by $\Gamma^+ S = 0$ and $X^+ = \tau$; then
the gauge-fixed Green-Schwarz action is given by
\begin{equation} \label{gswithD}
S = \int d^2 \sigma \left\{(\dot{X}^i)^2-({X^i}{'})^2 - \mu^2 (X^i)^2
- 2i \p_a X^\mu \bar{S} \Gamma_\mu (\delta^{ab} - \epsilon^{ab} \Gamma^{11})
        {\mathcal D}_b S \right\},
\end{equation}
where $a,b,\dots$ are worldsheet indices, ${\mathcal D}_b$ is the
pullback to the worldsheet of the supercovariant
derivative~\eqref{scdIIA},
and the overdot and prime respectively denote differentiation with respect to
$\tau$ and $\sigma$.
Explicitly,
\begin{multline} \label{gs}
S = \int d^2 \sigma \left\{(\dot{X}^i)^2-({X^i}{'})^2 - \mu^2 (X^i)^2 
\right. \\ \left.
-2 i \left[\bar{S} \Gamma^{-} (\p_\tau + \Gamma^{11} \p_\sigma) S
 - \frac{\mu}{4} \bar{S} \Gamma^- \Gamma^8 \Gamma^{11} S
 + \frac{1}{4} \bar{S} \Gamma^- \fs{\Theta} S \right] \right\}.
\end{multline}
The bosonic part of the action is the action of 8 massive bosons of
equal mass $\mu$; this is quite familiar by now~\cite{m,bmn,mt,rt,jm}.

The fermionic equation of motion is
\begin{equation} \label{gsferm}
(\p_\tau + \p_\sigma \Gamma^{11}) S - \frac{\mu}{4} \Gamma^8
\Gamma^{11} S + \frac{1}{4} \fs{\Theta} S = 0.
\end{equation}
Multiplying by $(\p_\tau -  \Gamma^{11} \p_\sigma)$ gives
\begin{equation}
(\p_\tau^2 - \p_\sigma^2) S - \frac{1}{16} (\mu \Gamma^8 \Gamma^{11} -
\fs{\Theta})^2 S = 0.
\end{equation}
Remarkably, it is precisely for $\Gamma^+ S=0$ that \hbox{$(\mu
\Gamma^8\Gamma^{11} - \fs{\Theta})^2 S = -16 \mu^2 S$}; thus
\begin{equation}
(-\p_\tau^2+\p_\sigma^2) S - \mu^2 S = 0,
\end{equation}
so the fermions have degenerate mass with the bosons, as required
given the supersymmetry.
[As an aside,
note that for an $S^+_{89}$ compactification, the only difference is
the sign of the Kaluza-Klein gauge field.  This compactification
preserves only the 16
supersymmetries annihilated by $\Gamma^+$, and so the gauge-fixed
action~\eqref{gs} then has no worldsheet (or dynamical) supersymmetries.
Indeed it is straightforward to see that the 16 physical fermions for
this less-symmetric background
have mass eigenvalues of $\frac{\mu}{2}$ (with degeneracy 10) and
$\frac{3\mu}{2}$ (with degeneracy 6).]

To quantize open strings on this background,
set
\begin{equation} \label{gsbc}
\evalat{S}{\text{boundary}} = M \evalat{S}{\text{boundary}},
\end{equation}
where $M$ is a matrix determined by the open string.  
Clearly $M^2=\one$, and 
since $M$ should relate the negative chirality part of $S$ to its
positive chirality part, it should anticommute with $\Gamma^{11}$.
Specifically,
\begin{equation} \label{defM}
M=\Gamma^{+-i_1i_2\dots i_{p-1}} \bigl(\Gamma^{11}\bigr)^{p/2+1},
\end{equation}
gives boundary conditions appropriate to
a D$p$-brane oriented parallel to the ``light-cone'' directions, and
$X^{i_1}\dots X^{i_{p-1}}$.  The factor of $\Gamma^{11}$ is
included for certain values of $p$ in order to ensure that $M^2=\one$.%
\footnote{For additional details on 
fermionic boundary conditions, see e.g.~\cite{lw}.
\skipthis{of course, there is an extra
$\Gamma^{11}$ in $M$, or equivalently the $\Gamma$-matrices in $M$
determine the directions transverse to the D-brane. 
}
The method described here was used to
find supersymmetric D-branes of the IIB maximally supersymmetric
pp-wave in~\cite{dp}.
}
Of course, since $M$ anticommutes with $\Gamma^{11}$, it must contain an
odd number of $\Gamma$-matrices, so $p$ is even for type IIA.

To find supersymmetric D-branes, we demand zero-modes of the fermions
$S$ by requiring that the boundary
condition~\eqref{gsbc} respect the
equation of motion~\eqref{gsferm}
when there are no excitations along the string:
$\p_\sigma S = 0$.  This results in the condition
\begin{equation} \label{fordbrane}
\com{M}{\mu \Gamma^{8} \Gamma^{11}-\fs{\Theta}} = 0.
\end{equation}
There do not appear to be any solutions of the form~\eqref{defM} to
this equation. 
It would be interesting, perhaps using techniques of~\cite{es}
or~\cite{sy,hs}%
, to
find the central extension to the algebra~\eqref{QQ}.
This would provide an additional proof of the absence of
supersymmetric D-branes in this background.
%Very recently, the central charges for the maximally supersymmetric
%pp-wave in M-theory were found using M(atrix) theory~\cite{hs}.

%\section{Other Solutions with at Least 26 Supercharges?}
\section{Discussion} 
\label{sec:unique}

We have presented an 11-dimensional pp-wave that
preserves 26 supercharges.  On compactification to 10 dimensions,
it gives a 26 supercharge IIA string that does not admit supersymmetric
D0-branes---or any supersymmetric D-branes.  Interestingly, for a different
compactification (on $S^+_{89}$ instead of $S^-_{89}$, say),
D0-brane charge {\em does\/} appear in the supersymmetry algebra, even
though that IIA theory only has 16 supercharges.

It is natural to wonder if the special pp-wave introduced here is in
any sense generic, and if there are solutions with 28 and 30
supercharges.  The pp-wave given in equation~\eqref{26pp} was found
via a general analysis of the eigenvalues $\rho_{(i)}$, with the aid of
\Mathematica.%
\footnote{The \Mathematica\
program and its output---which lists all
7-parameter solutions which preserve at least 20 supercharges---can be
found at \web{http://www.rci.rutgers.edu/~jmich/pp/all7.html} and
\web{http://schwinger.harvard.edu/~jeremy/pp/all7.html}.}
Although there are, of course, many solutions with 24 or
fewer supercharges, upon demanding at least 26 supercharges, the
analysis found only the
solution~\eqref{26pp}, and the maximally supersymmetric
solution~\cite{kg}, albeit in several coordinate systems.
Thus,% to the extent that the reader is willing to trust \Mathematica,%
%the reader will believe that 
\
the pp-wave~\eqref{26pp}, and the
pp-wave~\cite{kg}, are the only M-theoretic pp-waves that preserve at least 26
supercharges.
\skipthis{
I have also proven analytically that there are no solutions that
preserve 28 supercharges and are not the known maximally
supersymmetric solutions,
(but not yet
including $\lambda_1$ or $\lambda_2$ among the degenerate
eigenvalues).%
}% end skipthis and no space
In particular, there are no M-theory pp-waves
preserving 28 or 30 supercharges.

Note that the condition~\eqref{ns} is essential for this.
It is straightforward to satisfy the necessary but not sufficient
condition~\eqref{nis} for up to
six eigenvalues (or all eight).
For example, setting $n_1=n_2=n_3=0$ and
$n_5=n_6=n_7=-n_4=\mu$ sets
$\lambda_3=\lambda_4=\lambda_5=\lambda_6=\lambda_7=\lambda_8=0$.
However, this does not give a solution with 28 supercharges; for
example, $\rho^2_{3(2)} = 144\mu^2$ but $\rho^2_{5(2)}=0$ so the
necessary condition~\eqref{ns} is not satisfied.  Indeed, it is
straightforward, if tedious, to show that if
$\lambda_3=\lambda_4=\lambda_5=\lambda_6=\lambda_7=\lambda_8$ then the
condition~\eqref{nis} implies that the corresponding supersymmetric
solution
is either the maximally supersymmetric pp-wave~\cite{kg}, or
flat space, and thus has 32 supercharges, not 28.

It is, however, possible to find a 28 supercharge pp-wave in the type
IIB theory%
\skipthis{by, say, turning on the RR (or NS-NS) 3-form field strength
as $H= \mu dx^+ \wedge {^{(2)}}\Theta$, where ${^{(2)}}\Theta$ is a certain
symplectic 2-form with equal eigenvalues.  This leads
to a 28 supercharge pp-wave with U(4) rotational isometries---and for
which only the U(1) subgroup appears on the right-hand side of the supercharge
anticommutation relations.}%
.  One such solution and its superalgebra is given in the appendix.

\acknowledgments

I thank N. Lambert, H. Liu, and especially C. Hofman for enlightening
conversations.  I am also
grateful to J.~Gheerardyn for a helpful communication.
This research was supported in
part by DOE grant \hbox{\#DE-FG02-96ER40559}.

\appendix

\section{A 28 Supercharge IIB pp-wave} \label{sec:IIB}

In this section, we present a type IIB pp-wave and its 28
supercharge superalgebra.%
\footnote{This solution, and its $SL(2,\ZR)$ cousins,
was also recently presented in~\cite{br}.}
The field configuration is
\begin{subequations} \label{IIBsoln}
\begin{align}
ds^2 &= 2 dx^+ dx^- - \frac{1}{4} \mu^2 (x^i)^2 (dx^+)^2 + (dx^i)^2, 
i=1\dots8,
\\
^{(3)}F &= \mu dx^+ \wedge (dx^1dx^2 + dx^3 dx^4 - dx^5 dx^6 + dx^7 dx^8) 
     \equiv dx^+ \wedge {^{(2)}}\Theta,
\end{align}
\end{subequations}%
where for definiteness in the following, $^{(3)}F$ is the \RR\
field strength.  (Turning on the NS-NS field strength instead is equally good,
but modifies some equations below that involve spinors.)

The 28 Killing spinors are parameterized by a doublet of positive
chirality Majorana-Weyl spinors $\psi^\Lambda$, $\Lambda=1,2$, which
are preserved by the
projection operator
\begin{equation}
\proj = \frac{1}{16}
(\Gamma^{12}-\Gamma^{56}+\Gamma^{78})^2 \Gamma^{-} \Gamma^{+} +
\frac{9}{16} \Gamma^{-} \Gamma^+ + \frac{1}{2} \Gamma^+ \Gamma^-.
\end{equation}
Note that this rank 14 projection operator does not distinguish
between
or mix $\psi^1$
and $\psi^2$.  
The Killing spinors are
\begin{subequations}
\begin{equation}
\epsilon^\Lambda (\psi) = 
  \left(\one \delta_\Sigma^\Lambda - x^i  \Omega_i^\Lambda{_\Sigma}\right)
   \left(e^{\frac{1}{16} \fs{{^{(2)}}\Theta} (\Gamma^+ \Gamma^- + 2) 
       \rho_1}\right)^\Sigma{_\Pi} \psi^\Pi,
\end{equation}
where
\begin{equation}
\Omega_i^\Lambda{_\Sigma} = \frac{1}{16} \left[ 2 \fs{^{(2)}\Theta} \Gamma^i
  - \Gamma^i \fs{^{(2)}\Theta}\right] \Gamma^+
    \rho_1^\Lambda{_\Sigma},
\end{equation}
\end{subequations}%
and $\rho_1$ is the Pauli matrix \hbox{$\sigma^1=\begin{spmatrix} 0 &
1 \\ 1 & 0\end{spmatrix}$}.  The Killing spinors obey the differential
equation (see e.g.~\cite{gsw} and for recent applications to pp-waves, 
e.g.~\cite{clp2,khkw,bjlm})
\begin{equation}
\nabla_\mu \epsilon^\Lambda - \frac{1}{16} (2 \fs{{^{(3)}}F} \Gamma_\mu 
+ \Gamma_\mu \fs{{^{(3)}}F}) \rho_1^\Lambda{_\Sigma} \epsilon^\Sigma =
0,
\end{equation}
as well as the constraint, from the dilatino variation,
\begin{equation}
\fs{{^{(3)}} F} \epsilon^\Lambda = 0.
\end{equation}

Recognizing
that ${^{(2)}\Theta}$ is a simple symplectic form on flat $\ZR^8$, we
see that the
solution~\eqref{IIBsoln}
has rotational isometry group $\text{U}(4)\subset \SO(8)$.  In particular,
the central U(1) is \hbox{$\sorot = M_{12}+M_{34}-M_{56}+M_{78}$},
where \hbox{$M_{ij} = x^i \p_j - x^j \p_i$}.  Also, we have the usual
Heisenberg isometries
\begin{subequations}
\begin{align}
e_+ &= -\p_+, & e_- &= -\p_-, \\
e_i &= -\cos \frac{\mu}{2} x^+ \p_i - \frac{\mu}{2} x^i \sin \frac{\mu}{2}
x^+ \p_-, &
e_i^* &= -\frac{\mu}{2} \sin \frac{\mu}{2} x^+ \p_i + \frac{\mu^2}{4}
x^i \cos \frac{\mu}{2} x^+ \p_-.
\end{align}
\end{subequations}

In addition to the Killing spinors, $e_i$ and $e_i^*$,
transforming in the standard way
under the U$(4)\subset\SO(8)$ rotations, the supersymmetry algebra is
\begin{multline}
\bar{\epsilon}_\Lambda (\psi_1) \Gamma^\mu \epsilon^\Lambda(\psi_2) \p_\mu
= -(\bar{\psi}_{1\Lambda} \Gamma^{\ul-} \psi_2^\Lambda) e_-
- (\bar{\psi}_{1\Lambda} \Gamma^+ \psi_2^\Lambda) e_+
+ \frac{\mu}{2} (\bar{\psi}_{1\Lambda} \Gamma^+ \rho_1^\Lambda{_\Sigma} 
     \psi_2^\Sigma) \sorot \\
+ \frac{1}{2\mu^2} \left( \bar{\psi}_{1\Lambda}
     [ \fs{^{(2)}\Theta}\Gamma^i (\one + \half \Gamma^- \Gamma^+)
       -\Gamma^i \fs{^{(2)}\Theta}(\one + \half \Gamma^+ \Gamma^-)
     ] \rho_1^{\Lambda}{_\Sigma} \psi_2^\Sigma\right) e_i^*
- \bar{\psi}_{1\Lambda} \Gamma^i \psi_2 e_i,
\end{multline}
\begin{subequations}
\begin{align}
\lie{e_-}{\epsilon^\Lambda(\psi)} &= 0, \\
\lie{e_+}{\epsilon^\Lambda(\psi)} &= \epsilon^\Lambda
(-\frac{1}{16} \fs{^{(2)}\Theta} (\Gamma^+ \Gamma^- + 2) \rho_1 \psi), \\
\lie{e_i}{\epsilon^\Lambda(\psi)} &= \epsilon^\Lambda 
(\frac{1}{8} \fs{^{(2)}\Theta} \Gamma^i \Gamma^+ \rho_1 \psi), \\
\lie{e_i^*}{\epsilon^\Lambda(\psi)} &= \epsilon^\Lambda
(\frac{\mu^2}{8} \Gamma^i \Gamma^+ \psi),
\end{align}
\end{subequations}
\begin{subequations}
\begin{align}
\com{e_i}{e_j^*} &= \frac{\mu^2}{4} \delta_{ij} e_-, \\
\com{e_+}{e_i} &= e_i^*, \\
\com{e_+}{e_i^*} &= -\frac{\mu^2}{4} e_i.
\end{align}
\end{subequations}
In particular, note that the supergroup has a semidirect product
structure G$\rtimes$SU(4).

Also, note that all spacelike compactifications break at least some
supersymmetry.  This is in accord with the no-go
statement for 11-dimensional solutions with 28 supercharges.

\end{document}